%% file: qwcomp.tex
\documentclass[submission,copyright,creativecommons]{eptcs}
\providecommand{\event}{QSQW 2020} % Name of the event you are submitting to
\usepackage{underscore}           % Only needed if you use pdflatex.
\usepackage[mathcal]{euscript}  % redefs \mathcal to non-scr 
\usepackage{bbold}              % for nice identity symbol and others
\usepackage{amsmath}            % provides \text{} in math mode
\usepackage{subfigure}          % remove if not used
\usepackage{tikz}               % for figures
\usetikzlibrary{shapes,arrows}
\tikzstyle{every loop}=[]
\newcommand{\ie}{i.e.~}

\newcommand{\ket}[1]{\ensuremath{\left|#1\right\rangle}}
\newcommand{\bra}[1]{\ensuremath{\left\langle #1\right|}}
\newcommand{\p}{\mathbf{p}}
\newcommand{\s}{\mathbf{s}}
\newcommand{\RT}{\mathcal{R_T}}
\newcommand{\IRT}{\mathcal{\widetilde{R}_T}}
\newcommand{\theory}{\mathcal{T}}
\newcommand{\see}{C}
\title{How to Compute Using Quantum Walks}
\author{Viv Kendon
\institute{Physics Department\\Durham University\\Durham, UK}
\email{viv.kendon@durham.ac.uk}
}
\def\titlerunning{Computing with Quantum Walks}
\def\authorrunning{V.~Kendon}
\begin{document}
\maketitle

\begin{abstract}
Quantum walks are widely and successfully used to model diverse physical processes.
This leads to computation of the models, to explore their properties.
Quantum walks have also been shown to be universal for quantum computing.
This is a more subtle result than is often appreciated, since it applies to computations run on qubit-based quantum computers in the single walker case, and physical quantum walks in the multi-walker case (quantum cellular automata).
Nonetheless, quantum walks are powerful tools for quantum computing when correctly applied.
In this paper, I explain the relationship between quantum walks as models and quantum walks as computational tools, and give some examples of their application in both contexts.
\end{abstract}

%%%%%%%%%%%%%%%%%%%%%%%%%%%%%%%%%%%%%%%%%%%%%%%%%%%%%%%%%%%%%%%%%%%%%%%%%%%%%%
\section{Introduction}\label{sec:intro}

Quantum walks are widely appreciated for their mathematical intricacies and their many applications.
Examples abound, from transport properties 
\cite{Mohs2008,Boug2016}
to quantum algorithms
\cite{Shen2002,Chil2002}.
There is an important difference between how quantum walks are used for computing and how they are used for modelling physical systems.
For computing, obtaining an efficient algorithm is a key goal, whereas for models the goal is to accurately describe the physical characteristics of the system.
Quantum walk experiments
\cite{Pere2008,Kars2009,Broo2010,Schr2011}
have mostly implemented physical quantum walks, with a walker (photon, atom) traversing a path in the experimental apparatus as the quantum walk evolves.
There are a few implementations of quantum walks in an algorithmic setting, e.g.,
\cite{Ryan2005},
where the walk is encoded into qubits that label the position of the walker.
In this paper, I explain why this encoding step is critical for producing efficient quantum walk algorithms, and provide examples of algorithms that might be useful in the near future as quantum hardware develops.
The paper is organised as follows.  
In section \ref{sec:ART}, I outline a general framework for reasoning about physical theories, science, engineering, and computing.
This is followed in section \ref{sec:efficient} by discussion of what makes a computation ``efficient''.
Section \ref{sec:qwcompute} contains a discussion of how quantum walks are used for efficient algorithms in a continuous-time setting.
Section \ref{sec:conc} summarises the discussion and considers the future of quantum walk computing.

%%%%%%%%%%%%%%%%%%%%%%%%%%%%%%%%%%%%%%%%%%%%%%%%%%%%%%%%%%%%%%%%%%%%%%%%%%%%%%
\section{A framework for physical computing}\label{sec:ART}

%Define computing and science in the diagrams, and point out the necessity for an RE.
It is helpful to start by being more precise about what computing is.
A \emph{computation} is specified mathematically as a calcuation or process on input data.
While there is not a clear and precise definition, fuzzy definitions are sufficient for this paper, where we are interested in the possibilities and constraints that the laws of physics place on computation.
We can define a \emph{computer} as a physical system used to carry out a specific computation.
For example, a smart phone, your brain, an abacus -- very different physical systems can all be computers.
The origin of the term ``computer'' (up to about 1950) was humans doing calculations that were part of a larger overall calculation, so could be done faster in parallel.
Something we can immediately specify for all these different computers is that (successfully) carrying out an actual computation must produce an \emph{output}.
Any computation with no output can be replaced by a brick doing nothing.
This effectively rules out pancomputationalism, in which everything in the universe is considered to be computing \cite{Picc2017}.
This is not a useful stance to take, there is then no distinction between being a computer and being any other type of physical system. 
A tighter definition and framework is proposed in \cite{Hors2014}, outlined here as follows.

%--------------------------%
%% subsec 1
\subsection{Representation:}
%--------------------------%

\begin{figure}
\input{Figure1}
\caption{%
(a) Spaces of abstract and physical objects: here an electron $e^-$ and a quantum mechanical wavefunction $\psi$.
(b) The modelling representation relation $\mathcal{R}$ used to connect $e^-$ to $\psi$.
\label{fig:rep}
}
\end{figure}
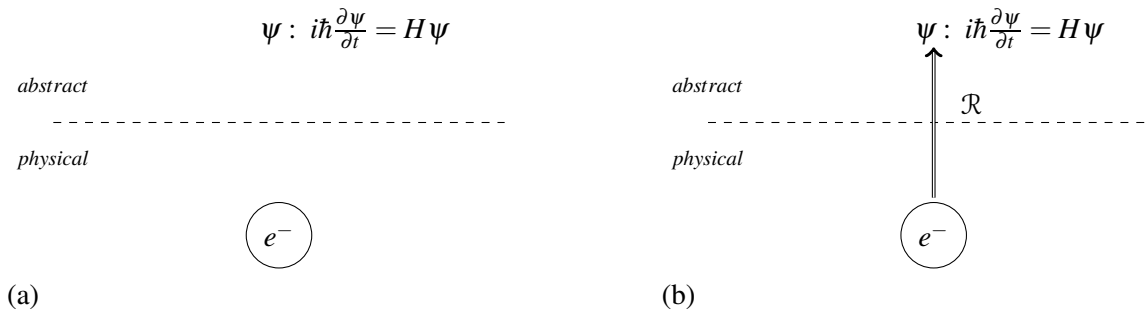

The basic building block of the framework is the relationship between a physical system and a mathematical description of it.  An example is shown in figure \ref{fig:rep}, of an electron $e^-$ and a wavefunction $\psi$ describing the electron.
%
%%%[Diagram of electron and wave function.]
These two different things, one physical, the other abstract, are related by what philosophers call the \emph{representation relation}
\cite{Putn1988,Hugh1997}, 
which we denote by $\mathcal{R}$.
Actually, there are several different representation relations. 
We are going to use two of them, \emph{modelling} and \emph{instantiation}.
In figure \ref{fig:rep}, we are using the representation relation $\mathcal{R}$ as the modelling relation, mediating from the \textit{physical} to the \textit{abstract}.
Importantly, note that representations are not unique. 
The quantum mechanical wavefunction is one useful choice, but we might alternatively choose to represent an electron as a classical trajectory in classical mechanics, or as a point charge in electrostatics.
The choice generally depends on what we happen to be interested in.
Thus, representations are theory dependent, and henceforth we will indicate this
by writing $\RT$ with the subscript $\theory$ indicating the relevant theory.
    
There are many philosophical open questions about representation.
We are going to leave the nuances of the open questions to the philosophers, and see what we can do with a framework based on abstract models of physical systems.
Despite the apparent separation between the abstract and physical realms in the diagrams, this is not a dualist theory.  
Everything in the diagrams is physical, everything abstract has to be instantiated in a physical representation, be that neuronal activity in your brain, squiggles on paper, electrons in a computer, or patterns on a display screen. 
This implies that there has to be a ``representational entity'', that owns the representation of the abstract model.  
This entity does not need to be human, or conscious, or necessarily even alive in the biological sense
\cite{Hors2017,Step2019}.
%This has consequences for defining computing, see section\ref{ssec:compute}.
Note that we are assuming some form of philosophical realism -- that there is a reality ``out there'' that we share and that persists when we don't observe it.  
This is a natural assumption for physicists, but if you don't agree with this, you arrive at different conclusions, which you are perfectly entitled to hold.

%----------------------------------%
%% subsec 2
\subsection{The scientific process:}\label{ssec:science}
%----------------------------------%

\begin{figure}
\input{Figure2}
\caption{%
(a) Physical system $\p$ is represented abstractly by $m_\p$ using the modelling representation relation $\RT$ of theory $\theory$.
(b) Abstract dynamics $\see_\theory (m_\p)$ give the evolved abstract state $m^\prime_\p $.
(c) Physical dynamics $\mathbf{H}(\p)$ give the final physical state $\p^\prime$.
(d) $\RT$ is used again to represent $\p^\prime$ as $m_{\p^\prime}$, the difference between $m^\prime_\p$ and $m_{\p^\prime}$ is denoted $\varepsilon$.
\label{fig:science}
}
\end{figure}
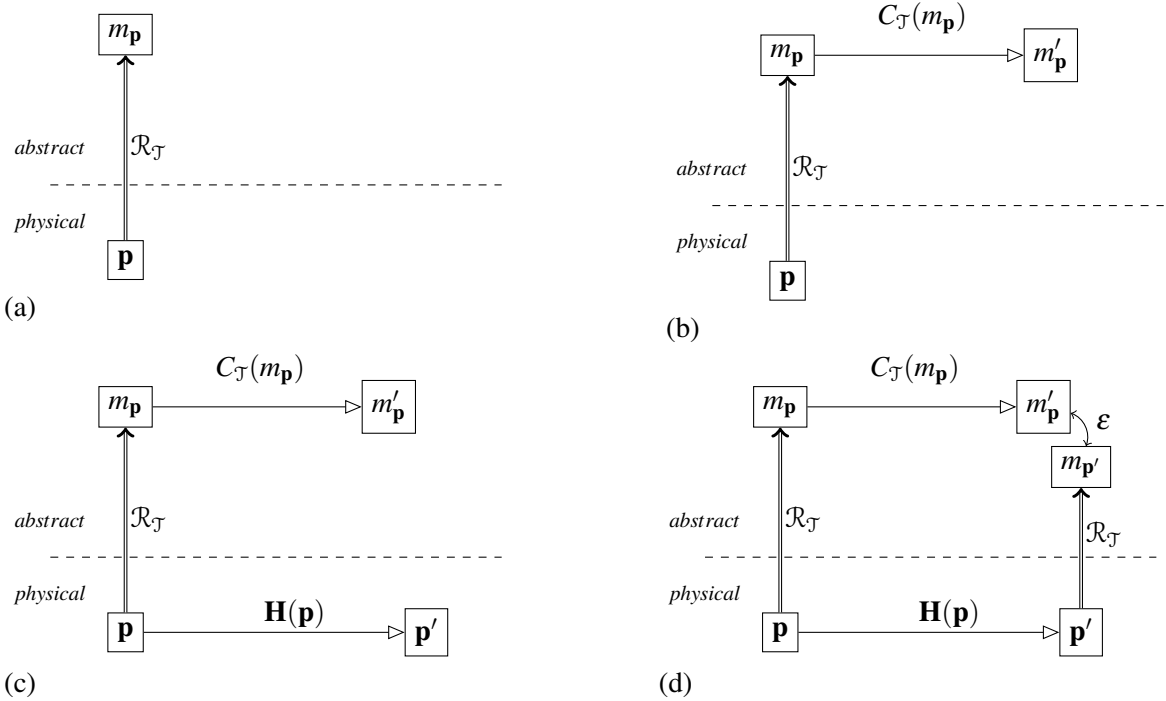

We can use the representation relation to describe how we do science, see figure \ref{fig:science}.
Given a physical object $\p$, we can use a theory $\theory$ to represent the object as $m_{\p}$, using the representation relation $\RT$, as in figure \ref{fig:science} (a).
Theory $\theory$ says that the object will evolve under some dynamics $\see_{\theory}$ to become $m'_{\p}$, as in figure \ref{fig:science} (b).
Experiment or observation of the physical object $\p$ as it evolves under its natural dynamics denoted $\mathbf{H}(\p)$ to become $\p'$ is shown in figure \ref{fig:science} (c).
We now wish to compare theory and experiment/observation.
To do this, we use the representation relation again, on $\p'$, to obtain the abstract object $m_{\p'}$.
We now have two objects of the same (mathematical) type, $m'_{\p}$ and $m_{\p'}$, that we can compare.
%
%%%[Draw/fill in diagram with scientific cycle and theory/experiment comparison.]
There are, of course, many reasons why we don't expect perfect agreement between experiment and theory.
The measurement precision in the experiment will be limited, the theory may not take into account all the possible errors, or we may know it is simpler than reality, but still a useful approximate theory.
We denote the difference between theory and experiment by $\varepsilon$, as indicated in figure \ref{fig:science} (d).
The goal of science is to develop theories that have a small enough $\varepsilon$ to be useful to make predictions; 
for a ``good'' theory, $m^\prime_{\mathbf{p}}\simeq m_{\mathbf{p}^\prime}$.
How we decide what makes a theory ``good enough'' is a major unsolved problem in the philosophy of science.
We will just assume that what we do works, based on good evidence of the effectiveness of science, despite philosophers not agreeing on exactly how.
Figure \ref{fig:science} is a simple building block in a much more complicated process.  
Science is a complex set of many interlocking diagrams like this, in which experimental equipment is repeatedly tested and characterised before being used for a new experiment.

%%%%%%%%%%
Prediction is the process of using our current theories to extrapolate into areas where we don't yet have experimental observations.  
Thus, we can design new experiments, based on predictions from our existing theories.
%
%%[Draw or outline part of cycle that gives prediction.]
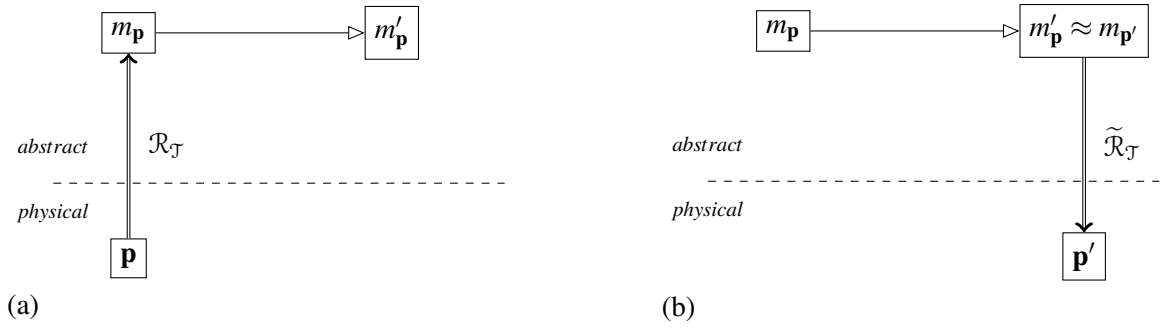
\begin{figure}
\input{Figure3}
\caption{%    
(a) The `predict process': abstract theory is used to predict physical evolution.
(b) The `instantiation process': using an instantiation representation relation $\IRT$, a physical object is found corresponding to the predicted evolution.
\label{fig:predict}
}
\end{figure}
Figure \ref{fig:predict} (a) shows how a good theory can be used to predict the outcomes of experiments that have not been done, including the possible existence of new particles or physical properties of systems.

Instantiation builds on prediction to locate or create an object predicted by theory, figure \ref{fig:predict} (b).
The instantiation representation relation $\IRT$ is in some sense the inverse of the modelling representation relation $\RT$, restricted to the class of objects that exist. 
It is possible to have theories of objects that do not actually exist, and it is through experiments that we discover whether or not a predicted object actually exists.

%------------------------------------------------%
%% subsec 3
\subsection{Engineering}
%------------------------------------------------%

\begin{figure}
\centering
\input{Figure4}
\caption{%
Making an artifact $p^{\prime}$ from a design specification $C_{\theory}(m_{\p})$ and raw material $\p$.
\label{fig:engineer}
}
\end{figure}
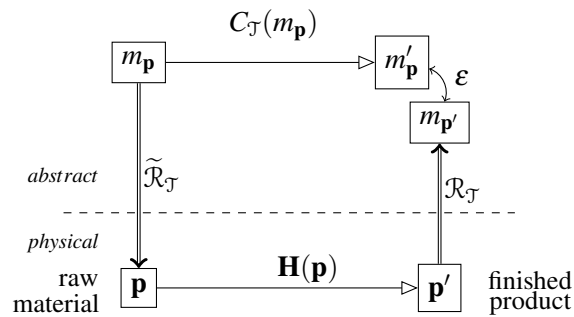

Once we have done enough science to develop a robust theory, in which our predictions and experimental results agree to within a small $\varepsilon$, we can put our theory to work.
Figure \ref{fig:engineer} shows how raw materials $\p$ are turned into finished products $\p^\prime$ following a design specification $C_{\theory}(m_\p)$.
The final step in the process is to check the quality of the production process by using the modelling representation relation $\RT$ on the finished product and comparing this with the design specification.
% Engineering figure
Engineering, i.e., building technology, effectively reverses the modelling representation relation to instantiate the physical objects $\p^{\prime}$ we want to make.
It is important to remember that $\IRT$ is a shorthand for a lot of science, to enable us to manufacture things reliably.

%%%%%%%%%%%%%%%%%%%%%%%%%%%%%%%%%%%%%%%%%%%%%
The difference between science and technology in terms of the abstraction/representation theory framework is that, if the abstract and physical don't match well enough, \ie, if $\varepsilon$ is too large, then for science we need to improve the theory $\theory$ until it describes our experiments well enough, but for technology we need to improve the production process for the physical system $\p'$ until it meets out design specification.

%------------------------------------------------%
%% subsec 4
\subsection{Computing}\label{ssec:compute}
%------------------------------------------------%

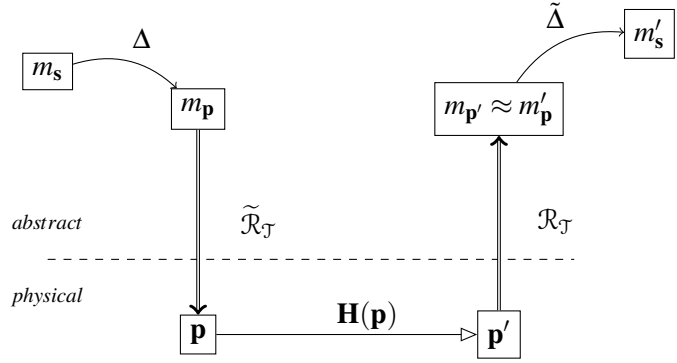
\begin{figure}
\centering
\input{Figure6}
\caption{%
Carrying out a computation on a physical computer $\p$.
The abstract problem $m_\s$ is first encoded $\Delta$ into the
model of the computer $m_\p$.
The computation is set up $\IRT$, and run, and the result read out $\RT$.
It is then decoded $\tilde{\Delta}$ into the solution $m^{\prime}_\s$.
\label{fig:compute}
}
\end{figure}

Among the many things we engineer are computers.
Figure \ref{fig:compute} illustrates the process of using a physical computer in terms of the framework.
The abstract problem we are trying to solve (labeled $m_\s$ in figure \ref{fig:compute}) must first be encoded into the abstract machine model $m_\p$ using encoding $\Delta$.
Then, we can instantiate $m_\p$ in $\p$ using the reverse representation relation $\IRT$.
The computation proceeds by evolving $\p$ into $\p^\prime$ using $\mathbf{H}(\p)$.
When it has finished, we measure or observe the output state $\p^\prime$, and represent it as model $m_\p^\prime$.
Now, we can assume $m_\p^\prime \simeq m^\prime_\p$, because the computer is based on sound science and has been engineered and tested to meet the design specification.
Finally, we translate $m^\prime_\p$ into $m^\prime_s$ using the appropriate decoding $\tilde{\Delta}$.
%
%%[Draw compute diagram, with encode and decode steps.]
The physical system $\p$ in figure \ref{fig:compute} is representing an unrelated abstract computation via the mathematical encode $\Delta$ and decode $\tilde{\Delta}$ steps.  
In practice, there may be many layers of abstract encoding (known as ``refinement'' in computer science), e.g., compiling the source code of a computer program \cite{Hors2014}.

The key property of representation, from a computational point of view, is that the physical representation is arbitrary. 
It could have used a different instantiation just as well, different computers can carry out the same computation, and different programs can be written that carry out the same computation. 
Trivially, a different physical representation is obtained by inverting all the zeros and ones in the binary encoding in conventional computers.  
Less trivially, the number two can be represented in many ways, including \{2, II, (ii), ($\bullet$ $\bullet$), two, ..\}.
Group theory makes this idea mathematically precise, distinguishing between the concept of a set with certain properties, like ``two-ness'', and a particular representation of that group like ($\bullet$ $\bullet$).
A rather different example, of bacteria using the same signalling molecule but with different meanings in different species, is explained in detail in
\cite{Hors2017}.

A further consequence of everything being physical is that a physical computation actually taking place includes a \emph{computational entity}, a representational entity that owns the computation and where the abstract representation is physically instantiated.  
Hence, it is clear that a brain is a computer (among other functions).
However, humans not are not required: a robot could use a calculator; a bacterium has no brain \cite{Hors2017}.
Those familiar with communications may have met Alice and Bob in presentations of communications theory or experiments.
Alice and Bob serve to remind us that communications involves the transmission of information between entities for which the information is meaningful in some way.
While we can do the theory entirely abstractly, what makes it communication rather than mathematics is the possibility of transmitting meaningful information between representational entities.

%---------------------%
%% subsec 5
\subsection{Simulation}
%---------------------%

%%[Draw simulation diagram.]
\begin{figure}
\input{Figure8} % simulation grid
\caption{%
(a) Predict process using $C_{\theory}(\p)$ to find the abstract prediction for the evolution of system $\p$.
(b) Encoding $C_{\theory}(\p)$ in the simulator model dynamics, $C_{\theory}(\mathbf{s})$ using $\Delta$ and $\tilde{\Delta}$.
(c) Adding the compute process: the physical simulator $\mathbf{s}$ now determines the abstract evolution $C_{\theory}(\mathbf{s})$ and via encode/decode $\Delta/\tilde{\Delta}$ calculates $C_{\theory}(\p)$.
(d) A system simulating itself, with encode and decode as identities.
Note that the models $m_\p$ and $n_\mathbf{s}$ must be present for this to be simulation.
\label{fig:fullsim}
}
\end{figure}
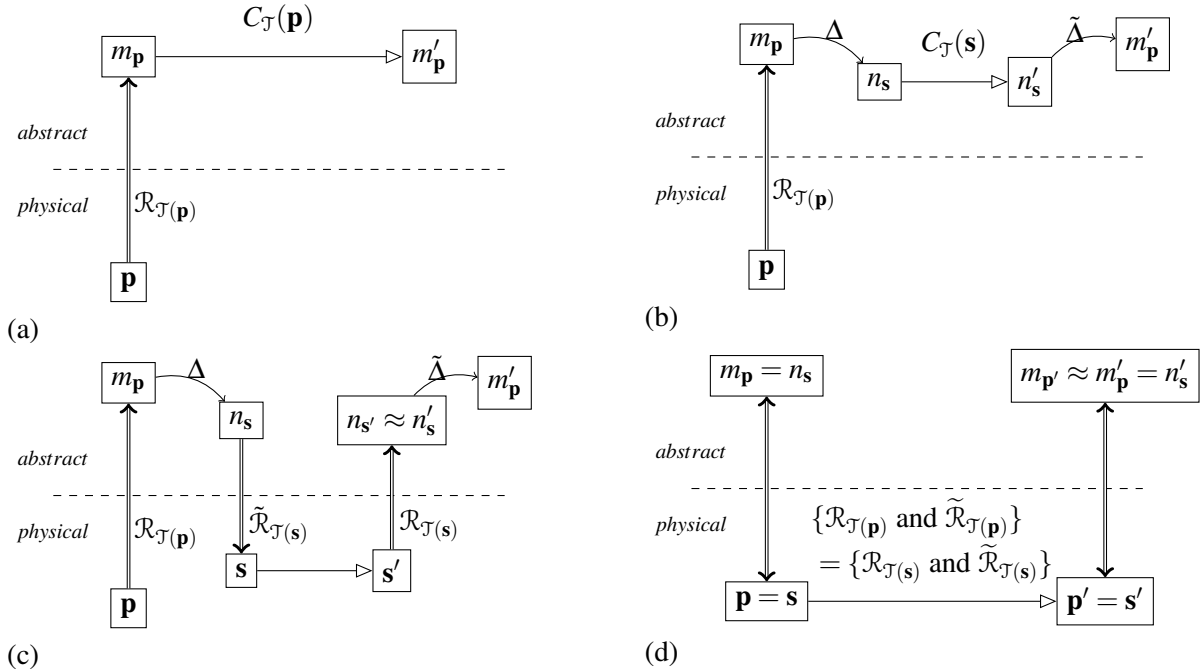
Simulation is calculating the details of a model of a physical system.
We can only simulate our models, it does not make sense to say we are directly simulating a physical system.  Working directly with a physical system is an experiment, or engineering.
Simulation is thus a particular type of computation that happens to be calculating something about the theory for a physical system.  
The diagrams for system $\mathbf{s}$ running as a simulator for system $\p$, constructed as a compute process nested within a predict process, are given in figure \ref{fig:fullsim}.
Figure \ref{fig:fullsim} (a) is the normal prediction process, where a model $m_\p$ is used to predict the properties of $\p$.
Figure \ref{fig:fullsim} (b) is an intermediate step, where the model of the simulator (computer) $m_\mathbf{s}$ is used to set up the encode $\Delta$ and decode $\tilde{\Delta}$ steps, and the program $C_{\theory}(\mathbf{s})$ to run the simulation.
Figure \ref{fig:fullsim} (c) shows the simulation taking place, with the program $C_{\theory}(\mathbf{s})$ running on the physical simulator $\mathbf{s}$.
Note that the theories in $C_{\theory}(\mathbf{s})$ and $C_{\theory}(\mathbf{p})$ do not need to be the same.  They have been shown as the same in figure \ref{fig:fullsim} to avoid cluttering the notation.
A computer can simulate itself, see figure \ref{fig:fullsim} (d), and this is often done, as \emph{virtual machines}.  In figure \ref{fig:fullsim} (d), the self-simulation is shown with the encode and decode as the identity.  For virtual machines, the encode and decode are usually not the identity, with the physical computer simulating one (or more than one) approximate copy of itself.
However, most physical systems do not simulate themselves in real time.
Unless there is representation, there is no theory for what is being calculated, and no computational or representational entity for which the computation is meaningful.
Computation as defined in this framework is a high level process that must involve representation
\cite{Hors2014,Hors2017}.

%%%%%%%%%%%%%%%%%%%%%%%%%%%%%%%%%%%%%%%%%%%%%%%%%%%%%%%%%%%%%%%%%%%%%%%%%%%%%%
\section{Efficient computation}\label{sec:efficient}
%-----------------------------%

In the abstraction/representation theory framework just outlined, computing is defined as a high level process that involves engineered (or evolved) objects and computational entities.  
The requirement for a computational entity implies computing needs to be purposeful, or useful.  
Which implies that, in general, computing is done in order to efficiently solve a problem.
We now turn to the question of what is needed for efficient computing, what it is that makes one computer better than another for a particular computation.  
There are many facets to this question and we will only consider a few key points that are important for quantum computing.

The first important consideration is that \emph{efficient} has different criteria for different purposes.  
On the one hand, complexity theorists consider asymptotic scaling of computations.  For them, efficient is usually defined as a polynomial scaling of the computational resources with respect to the problem size.
On the other hand, for practical computation, \emph{efficient} means obtaining answers on the timescales relevant for the computational entity.
For rendering graphics, the timescale is faster than the eye detects flickers.
For hard simulations, days or weeks are acceptable.
For real-time monitoring or prediction, it means faster than the physical process being modelled: an accurate weather forecast is not nearly so useful the day after the weather has already happened.
For quantum computing, the complexity of a problem and the predicted quantum advantage are a good starting point, but not the whole story when it comes to building a useful quantum computer.
A more favourable prefactor in the scaling can be enough to make using a quantum computer worthwhile, especially for time-sensitive computation. 
It isn't necessary to solve a larger problem than is possible classically to gain an advantage.
Solving a problem faster can be enough to gain a competitive advantage, depending on the value of the solution.

\begin{table}\label{tab:unary}
\caption{%
Relationship between unary and binary encoding of numbers.
}
\centering
\begin{tabular*}{0.7\textwidth}{r@{\extracolsep{\fill}}r@{}r}
\textit{Number} & \textit{Unary} & \textit{Binary}\\
\hline
0 &    & 0 \\
1 & $\bullet$ & 1 \\
2 & $\bullet\bullet$ & 10\\
3 & $\bullet\bullet\bullet$ & 11\\
4 & $\bullet\bullet\bullet\bullet$ & 100\\
$\cdots$ & $\cdots$ & $\cdots$ \\
8 & $\bullet\bullet\bullet\bullet$ $\bullet\bullet\bullet\bullet$ & 1000\\
$\cdots$ & $\cdots$ & $\cdots$ \\
$N$ & $N \times \bullet$ & $\log_2N$ bits \\
$\cdots$ & $\cdots$ & $\cdots$ \\
example & database records & index of records \\
\end{tabular*}
\end{table}
A second important consideration is efficient encoding of the data.  
Binary encoding is one reason why classical digital computers work so efficiently.
Table \ref{tab:unary} shows how binary encoding is exponentially more efficient in terms of memory use compared with unary.
This becomes increasingly important as the size of the numbers increases.
Modern digital computers use even more efficient representations, such as floating point numbers, which provide a trade off between efficiency and precision.

The same encoding considerations apply when building quantum computers
\cite{Eker1998,Blum2002}.
In addition, because the measurement process in quantum mechanics is non-trivial and can alter the state of the register, the encoding has an impact on the efficiency and accuracy of the measurement process.
For a unary encoding, it is necessary to measure in a way that can distinguish between $N$ different outcomes.
This becomes increasingly challenging as $N$ becomes large.
For binary encoding, $\log_2N$ measurements with two outcomes each is enough to distinguish $N$ numbers.  This is more efficient and better for accuracy, since each measurement only has to distinguish between two orthogonal outcomes.
Doubling the size of the problem ($N$) only requires one more binary measurement, which is more efficient than doubling the number of measurement outcomes that must be distinguished.
%For example, compare the task of identifying a single pixel in an image of $N$ pixels with a measurement process that identifies which half of the image the pixel is in, then which half of that half, and so on until the pixel is uniquely identified.
%The one-shot process involves resolving all the pixels at once, while the iterative process eliminates groups of pixels at a course-grained level, allowing the measurements to be refined as the number of pixels left under consideration reduces.
%Also, it is easier to recover from an error. 
%If the wrong half is selected, this will be identified at the next step and the other half can be selected instead to continue, at a cost of one extra binary measurement.
%An error in the full image setting will require the whole image to be reanalysed.

\begin{figure}
\input{hypercubefig}
\caption{%
Hypercubes and qubits for $n=1$, 2, 3, showing the encoding from quantum walk graph (upper) of size $N=2^n$ to qubit graph (lower) of size $n$.
The red dotted lines indicate possible interaction terms between qubits, that can be used to encode problems into Hamiltonians.
\label{fig:hypercubes}
}
\end{figure}
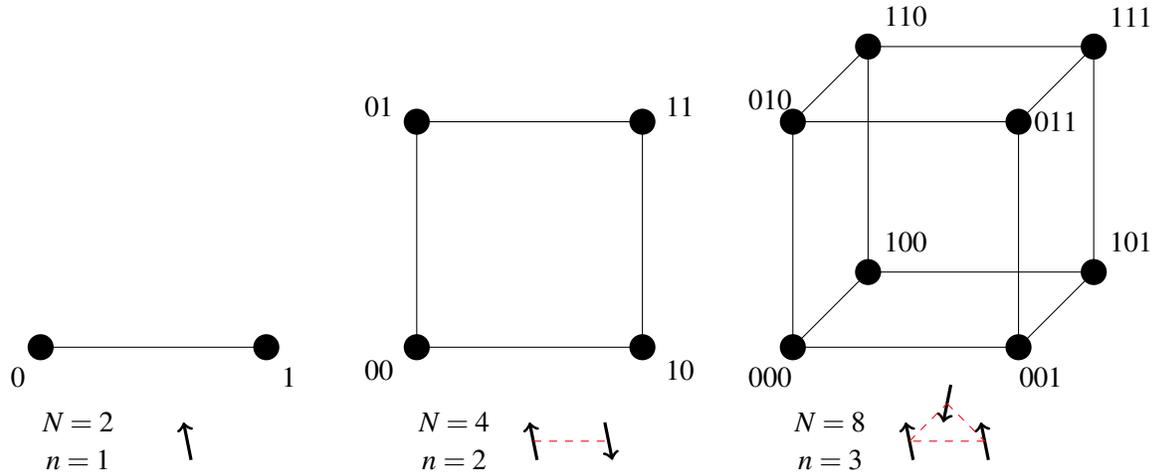

In a quantum walk setting, a useful example of the efficiency of encoding is a quantum walk on a hypercube, see figure \ref{fig:hypercubes}.
A quantum walk on the vertices of an $n$-dimensional hypercube can be encoded into $n$ qubits.
An $n$-dimensional hypercube has $N=2^n$ vertices, which are labelled by the bitstrings of the qubit basis states, as shown in figure \ref{fig:hypercubes}.
The relationship between the graph in Hilbert space representing the graph on which the quantum walk takes place, and the graph formed by the qubits storing the labels of the vertices, which exist in real physical space, is the same as the relationship between unary and binary encoding in table \ref{tab:unary}.
Quantum walks on hypercube graphs are especially useful for quantum computing, as will be explained in the next section.
The efficiency of encoding into qubits is not just for hypercube graphs.
Since any graph can have its vertices labeled with binary numbers, efficiently computing the evolution of the quantum walk on the graph can be done using the labels encoded into qubits.
A physical quantum walk is not an efficient way to compute how the quantum walk evolves.
This is easily illustrated by noting the ease of computing thousands or millions of steps of a quantum walk on a line using a classical digital computer.
This is far beyond any physical quantum walk experiments currently realised.  
However, fifty or more logical qubits would be enough to calculate a quantum walk beyond the capabilities of classical computers.
Fifty qubits is not far beyond the capacity of current quantum computers from IBM or Google, although they are currently too noisy to carry out a long enough quantum computation.

%%%%%%%%%%%%%%%%%%%%%%%%%%%%%%%%%%%%%%%%%%%%%%%%%%%%%%%%%%%%%%%%%%%%%%%%%%%%%%
\section{Quantum computing with quantum walks and related models}\label{sec:qwcompute}
%-------------------%

Important results from Childs et al.~%
\cite{Chil2009,Love2010},
proved that quantum walks are universal for quantum computing.
However, this result for single walkers applies strictly only to quantum walks encoded into qubits.
Without this encoding, the quantum walks are not efficient in the sense of complexity theory scaling with problem size, and thus cannot provide a quantum advantage.
A second result from Childs et al.~%
\cite{Chil2013}
proving that multiple interacting quantum walkers provide universal quantum computing does not require encoding into qubits.
In this case, the physical quantum walks are efficient, equivalent to quantum cellular automata \cite{Wies2008}.
Boson sampling \cite{Aaro2010} is halfway betweem the two, consisting of multiple non-interacting quantum walkers.
Boson sampling does not require encoding into qubits, but it is not universal for quantum computing.
It can still solve certain hard problems, such as approximating the permanent of a complex matrix, more efficiently than the best known classical algorithm.

There are several quantum walk algorithms with complexity proofs of their speed up over classical algorithms for the same problem.
Searching in an unsorted database is the first such algorithm
\cite{Shen2002}.
Quantum walk search is now a workhorse subroutine for theorists constructing algorithms for more complex problems,
providing a quadratic speed up over classical.  
For an unstructured database, random guessing in the best classical strategy available, which scales linearly with the problem size $N$, while quantum search scales as $\sqrt{N}$.
The ``glued trees'' algorithm 
\cite{Chil2002}, 
provides an exponential speed up with respect to an oracle, but does not lead to applications in the same way as quantum search.
Both search and the glued trees algorithm are permutation symmetric problems.
Permutation symmetric problems are useful for testbed algorithms, to test the performance of quantum hardware.  
Problems which can be solved analytically, efficiently computed classically (assuming some knowledge of the solution), and implemented efficiently on a  quantum computer, support detailed testing of the performance of quantum hardware.  
The implementation on quantum hardware can be efficiently achieved using extra qubits configured as gadgets to enforce the permutation symmetry 
\cite{Dodd2019}.

%-----------------------------------------------%
\subsection{Quantum computing in continuous time}
%-----------------------------------------------%

We can conveniently represent quantum computing in a flattened version of the AR diagram,
\begin{center}
input$\longrightarrow$\fbox{encode}$\longrightarrow\ket{\psi_{in}}\longrightarrow \hat{U} \longrightarrow\ket{\psi_{out}}\longrightarrow$\fbox{decode}$\longrightarrow$result,
\end{center}
which shows the steps to encode a problem into a quantum computer, and carry out the computation by evolving the quantum state $\ket{\psi_{in}}$ using unitary operation $\hat{U}$ to produce $\ket{\psi_{out}}$.
More generally, $\hat{U}$ can include interaction with the environment; this can also be modelled as unitary interactions in a larger Hilbert space that includes the environment.
The unitary transformation $\hat{U}$ can be done by a sequence of quantum gates in the circuit model.
A different approach is to engineer a Hamiltonian $\hat{H}(t)$ such that 
\begin{equation}
\ket{\psi_{out}} = \exp\{-i\hbar \int\!\!dt\,\,\hat{H}(t) \} \ket{\psi_{in}}
\end{equation}
The motivation for considering this form of processing for quantum computing is as follows.
Classical bits are either zero or one, and can flip between the two values.
On the other hand, qubits can be any superposition: $\alpha \ket{0} + \beta \ket{1}$, and can change smoothly from zero to one or anything in between.
Hence, discrete gates make sense for bits: bit-flip is all you can do.
But for qubits, exact bit-flip is as hard as other rotations.
Processing in continuous-time evolution thus makes sense for qubits.
This is related to work on the foundations of quantum mechanics by Lucien Hardy 
\cite{Hard2001}.
The axiom that distinguishes quantum from classical in a discrete state space is the possibility of evolving smoothly from one state to another, instead of undergoing a discontinuous change.

There are several ways to quantum compute in continuous time: quantum walks; adiabatic quantum computing; quantum annealing; and special purpose quantum simulation.
Each is described briefly in turn in the following subsections, apart from quantum simulation, for which see 
\cite{Brow2010} 
for a detailed review.

%-----------------------------------%
%% subsec 
\subsection{Computing by quantum walk}
%-----------------------------------%

First introduced in the continuous-time form by Farhi and Gutmann
\cite{Farh1998}, 
and used for an algorithm with exponential speed up by Childs et al.~%
\cite{Chil2002}.
Continuous-time quantum walks take place in a Hilbert space corresponding to a graph $G$, with basis states $\ket{j}$ for the label $j$ of each vertex of the graph.
If the graph has adjacency matrix $\mathbf{A}$, defined as $A_{jk}=1$ iff $\exists$ an edge between vertices $j$ and $k$, the Laplacian of the graph is defined by $\mathbf{L} = \mathbf{A} - \mathbf{D}$, where $\mathbf{D}$ is diagonal matrix $\mathbf{D}_{jj} = deg(j)$,
\ie the degree of vertex $j$ of the graph.
For undirected graphs, $\mathbf{L}$ is a symmetric matrix, and can be used to define the quantum walk Hamiltonian $\hat{H}_G$ such that $\langle j|\hat{H}_G|k\rangle = -\gamma \mathbf{L}_{jk}$, where $\gamma$ is the transition rate (probability of moving to a connected vertex per unit time).
The quantum walk evolves the initial state $\ket{\psi(0)}$ according to the Schr\"odinger equation, giving
\begin{equation}
\ket{\psi(t)}=\exp\{-i\hat{H}_Gt\}\ket{\psi(0)}.
\end{equation}
By itself, this doesn't provide an algorithm.
There are two ways to obtain a quantum algorithm from this type of quantum walk.
One way is to use a specific graph that defines the problem.  
This is the approach taken in the ``glued trees'' algorithm
\cite{Chil2002},
where the problem is to find the exit node from the entrance node (the roots of the trees).
The second way is to combine the quantum walk Hamiltonian with another Hamiltonian that specifies the problem.
This is the approach taken in the search algorithm
\cite{Shen2002},
where the marked state is treated differently from the rest of the vertices in the graph.
In all cases, the quantum walk must be encoded into qubits, as explained in section \ref{sec:efficient}, in order to obtain an efficient quantum algorithm.
As an example of this encoding, the quantum walk on a hypercube graph of size $n$ has $N=2^n$ vertices, but can be computed using $n$ qubits.
The Hamiltonian is given by
\begin{equation}\label{eq:hypercube}
\hat{H}_h = \gamma\left(n\mathbb{1} - \sum_{j=1}^n\sigma_j^{(x)}\right),
\end{equation}
where $\sigma_j^{(x)}$ is the Pauli-$x$ operator acting on the $j$th qubit.
The $\sigma_j^{(x)}$ operators flip the $j$th qubit, equivalent to walking along the edges in the $j$th direction on the hypercube, see figure \ref{fig:hypercubes}.

%---------------------%
\subsection{Adiabatic quantum computing}
%---------------------%

Introduced also by Farhi and Gutmann
\cite{Farh2000}
as natural way to solve optimisation problems such as boolean satisfiability (SAT) problems. 
The idea is to encode the solution into the ground state of a quantum Hamiltonian 
$\hat{H}_{\text{problem}}$.
Methods are known to do this efficiently
\cite{Choi2010,Lode2019}.
Then, starting in an easy-to-prepare Hamiltonian ground state of $\hat{H}_{\text{simple}}$, evolve the Hamiltonian from $\hat{H}_{\text{simple}}$ to $\hat{H}_{\text{problem}}$ smoothly,
\begin{equation}
\hat{H}_{AQC}(t) = \{1-s(t)\}\hat{H}_{\text{simple}} + s(t)\hat{H}_{\text{problem}}
\end{equation}
with $s(0) = 0$, $s(t_{f}) = 1$. 
If the time evolution is slow enough, the adiabatic theorem states that the quantum system will stay in the instantaneous eigenstate -- in this case the ground state -- throughout the time evolution with high probability.
The key question is thus, how slow is slow enough.   
The speed is in general determined by the minimum gap between ground state and excited states, which you don't know, because calculating it is equivalent to solving the problem.  
There are many subtleties to how slow is ``slow enough'', but since we design the problem Hamiltonian, we can avoid the more pathological cases where the adiabatic theorem fails.

Adiabatic quantum computing as been controversial, since it looks like can solve NP-hard problems, and it would be surprising if it could do this efficiently from what we know about complexity theory.
However, it is now clear it is likely to require exponential time, so would not be ``efficient'' according to theoretical CS.
Nonetheless, any speed up is potentially useful in practice, so adiabatic quantum computing is still very much of interest in a practical setting.
Not least, this is because it has been proved to be equivalent to digital quantum computation, that is, it can solve all the same problems using equivalent
resources
\cite{Ahar2004,Kemp2004}.
It is also potentially more resistent to some errors 
\cite{Chil2001},
although the scaling and propagation of errors is not known in general in this setting. 

Testbed implementation of adiabatic quantum computing has been done in NMR
\cite{Steff2003},  
but larger scale devices tend to be too noisy currently to be effective for adiabatic quantum computing.
Introduced by Finilla et al,
\cite{Fini1994},
quantum annealing is a noisy version of adiabatic quantum computing, with a low temperature bath that helps to remove energy to reach the ground state solution.
Quantum tunneling helps to avoid the annealing process becoming stuck in false minima.
The transverse Ising model is a particularly natural setting for quantum annealing
\cite{Kado1998}.
Note that in the literature, the terms \emph{adiabatic quantum computing} and \emph{quantum annealing} are used interchangeably, which can lead to confusion: they are in fact two different computational models.
While adiabatic quantum computing is what D-Wave Systems Inc.~would like to be doing,  what they are actually doing is closer to quantum annealing, with a lot of other types of unwanted noise, as well as a cold environment assisting the computation.

%---------------------%
\subsection{Hybrid QW-AQC algorithms}\label{ssec:hybrid}
%---------------------%

The reason for mentioning adiabatic quantum computing in a paper about quantum walks will become clear with an example.
Quantum search algorithms are well-studied and well-understood in both discrete (gate model, quantum walks) and continuous-time settings.
The search problem is to find the marked state in an unsorted database.
In a continuous-time setting, the problem Hamiltonian is simple to define,
\begin{equation}
\hat{H}_p = \hat{H}_m = \mathbb{1} - \ket{m}\bra{m}.
\end{equation}
This works by making the marked state $\ket{m}$ one unit of energy lower than the rest of the states.
In terms of Pauli operators on qubits, this does not look so simple,
\begin{equation}
\hat{H}_m = \mathbb{1} - \frac{1}{2^{n}}\prod_{j=1}^{n}(\mathbb{1} + q_j\sigma^{(z)}_j),
\end{equation}
where $q_j \in\{-1,1\}$ defines the bitstring corresponding to $m$ for $-1\equiv 0$ to convert to bits, and $\sigma^{(z)}_j$ is the Pauli-$z$ operator acting on the $j$th qubit.
There are techniques using extra qubits and perturbation theory that can implement this and other permutation symmetric problem Hamiltonians
\cite{Dodd2019}.

To implement a continuous-time quantum search algorithm, the full Hamiltonian is
\begin{equation}
\hat{H}_{QWS} = \hat{H}_G + \hat{H}_m,
\end{equation}
where $\hat{H}_G$ is the quantum walk Hamiltonian on graph $G$.
If the hypercube graph is used, $\hat{H}_G = \hat{H}_h$, as given in equation (\ref{eq:hypercube}).

The initial state of the qubits is
\begin{equation}\label{eq:equal}
\ket{\psi(t=0)} = 2^{-n/2}\sum_{j=0}^{2^n-1}\ket{j} = 2^{-n/2}(\ket{0}+\ket{1})^{\otimes n}
\end{equation}
which is the superposition of all possible states, and also happens to be the ground state of $\hat{H}_h$.
This is a natural choice for an unbiased initial state, it faithfully represents our lack of knowledge of the identity of the marked state.
Applying the quantum walk search Hamiltonian $\hat{H}_{QWS}$ to the initial state for a time $t_f\simeq\pi\sqrt{N}/2$ results in a final state $\ket{\psi(t_f)}\simeq\ket{m}$. 
A measurement of the qubits produces the index $m$ of the marked state with high probability, thus solving the search problem in time $O(\sqrt{N})$, a quadratic improvement over the classical best strategy.

The adiabatic quantum search algorithm looks quite similar to the quantum walk search algorithm,
\begin{equation}\label{eq:hybrid}
\hat{H}_{AQC}(t) = A(t)\hat{H}_h + B(t)\hat{H}_m
\end{equation}
where $A(t)$ and $B(t)$ specify the time dependent proportions of the two component Hamiltonians, and satisfy $A(t_f) = B(0) = 0$, $A(0) = B(t_f) = 1$ and $0\le A(t),B(t)\le 1$ for $0\le t \le t_f$.
As for the quantum walk, the qubits start in the initial state in equation (\ref{eq:equal}), the ground state of $\hat{H}_h$, and are measured after a time $t_f \propto \sqrt{N}$ to obtain a quadratic quantum speed up.

Both quantum walk and adiabatic quantum search algorithms are analytically solvable
\cite{Chil2004,Rola2002}
in the large $N$ limit, because they reduce to a two state single avoided crossing model 
\cite{Chil2004,Morl2019}.
There are many subtleties involved in setting the parameters $A(t)$, $B(t)$, $\gamma$, to achieve a quantum speed up, for detailed discussion and results, see 
\cite{Morl2019}.

The important observation for this paper is that quantum walk search and adiabatic quantum search are special cases of the same algorithm, with different choices of $A(t)$ and $B(t)$ in equation (\ref{eq:hybrid}) giving either quantum walk or adiabatic.
The computational mechanisms are different in each case.
For quantum walk, the system oscillates between the equal superposition initial state, and the marked state, while for adiabatic, the system transitions steadily from the initial state to the marked state.
A smooth interpolation between adiabatic and quantum walk can be done, and optimal contributions from both mechanisms employed to improve performance on imperfect hardware
\cite{Morl2019}.

Identifying the connections between quantum walk computing and adiabatic quantum computing allows several useful generalisations to be explored that are suitable for implementation on early quantum computers.
Any algorithm that has been shown to work in an adiabatic setting is likely to be promising in a quantum walk setting, too.
An example is finding the ground state of Sherrington-Kirkpatric spin glasses, for which adiabatic quantum computing is known to compare well against (classical) simulated annealing
\cite{Kado1998}.
Callison et al \cite{Call2019} recently showed that quantum walks can find the ground state better than quantum search algorithms, using many repeated short runs.
Quantum walks are potentially easier to implement experimentally, given their time-independent Hamiltonians.  
They are effective when used with many short runs and repeats, which suits early, noisy quantum hardware.
The benefits of combining both quantum walk and adiabatic techniques can be demonstrated in a range of settings, especially in the presence of hardware imperfections
\cite{Morl2019}.
However, it is important to use quantum algorithms where they will be most effective, to obtain a practical quantum advantage.

%---------------------%
\subsection{Hybrid classical-quantum algorithms}
%---------------------%

A quantum speed up using a quantum walk algorithm does not necessarily mean a quantum advantage overall for the computational task.
For search in an unsorted database, quadratic is the best possible speed up, and has been proven to give an advantage over the bext possible classical algorithm
\cite{Benn1997}.
However, the best classical algorithm is likely to be significantly better than random guessing, if the problem has some structure or correlations that can be exploited.
Real world problems always have such correlations, this is what makes them interesting and non-trivial to solve.
In the case of the spin glass ground state problem mentioned in section \ref{ssec:hybrid}, the best known classical algorithm 
\cite{Hart1984} 
is actually better than the pure quantum walk algorithm of
\cite{Call2019}.
However, a hybrid discrete-time quantum walk-classical algorithm that beats both has been presented by Montanaro
\cite{Mont2019}.
This is an example of one route to obtaining a quantum advantage.
Starting with the best classical algorithm, locate the bottleneck subroutine, and if amenable to a quantum speed up, replace it by a quantum version of the subroutine.
Another example also from Montanaro 
\cite{Mont2015}
applies this technique in the setting of backtracking algorithms.
Quantum walk search subroutines provide a further quadratic speed up compared with the classical algorithm.
More examples of how to construct hybrid algorithms, in a continuous-time setting, can be found in 
\cite{Chan2016,Chan2017}.

Employing quantum computers for the bottenecks in classical algorithms fits naturally into the diversification that is taking place currently in computer hardware.
Since we have reached the limits of standard silicon-based computer chips in terms of their speed of operation,
% \cite{Moor2009},%
computer chips have become more specialised, to speed up computation of commonly used tasks.
Graphics co-processors are one of the most widespread examples.  
They started out efficiently updating the screen image in real time, and graduated to computational science as massively parallel chips using shared memory.
Today's computers also typically include a dedicated chip for communications tasks such as ethernet and wifi.
A typical high performance computing facility now includes GPUs and field programmable gate arrays (FPGAs) as well as many conventional CPU cores.  
A quantum co-processor is a natural extension of this trend, and will enable hybrid quantum-classical algorithms to be run on well-integrated hardware.

%%%%%%%%%%%%%%%%%%%%%%%%%%%%%%%%%%%%%%%%%%%%%%%%%%%%%%%%%%%%%%%%%%%%%%%%%%%%%%
\section{Outlook}\label{sec:conc}
%---------------%

%Recap key ART messages of how QW as model relates to QW as computational tool.

Quantum walks are widely used as models for physical processes, especially transport properties.
They are also a useful tool for quantum algorithms.
It is important to understand the different requirements for different uses of quantum walks.  
Mathematical models of physical processes are abstract constructions, their usefulness depends on comparision with experiments on physical systems, as elaborated in section \ref{ssec:science}.
If the model does not describe the experimental results well enough, the model needs to be improved.
These mathematical models can then be explored using computation, especially if the analytical solutions are complicated or intractable.
A simple quantum walk on the line is easy to compute classically for large numbers of steps (thousands or millions).
As quantum walks spread ballistically (linearly in time), the computational resources required grow linearly, and this scaling is generally considered to be efficient from a computational point of view.
For quantum walks used as a computational tool, different criteria apply for when we might gain a computational advantage from quantum over classical.  
Efficient computation demands that the computation is encoded efficiently, and for quantum walks with single walkers, this means binary encoding into qubits (or equivalent) is essential.
A physical quantum walk with a single walker cannot provide efficient quantum computing.
Nonetheless, used correctly, quantum walks are a powerful computational tool.

Continuous-time quantum walks are part of a family of related continuous-time computational models, including:
adiabatic quantum computing; 
quantum walk; 
quantum annealing; 
special purpose quantum simulation.
These computational models all
use qubits for efficient binary encoding with continuous-time evolution to exploit the natural time-evolution of quantum mechanics.
The addition of cooling through a low temperature environment can offset some of the damaging effects of noise on the quantum computation.
Relating all these different computational models allows us to try different models and combinations to find the best for each particular problem.
Especially interesting is to try quantum walks on problems so far only examined in an adiabatic quantum computing setting.

There are a few caveats to be aware of.
Increasing numbers of qubits require more precise Hamiltonian parameters to be specified in the controls for the quantum computer, e.g., the coupling strengths between pairs of qubits in the problem Hamiltonian.
There is evidence from D-Wave that they are reaching that limit at around 2000 qubits.
However, 2000 qubits is still far more than classical computers can simulate, and the limit could be higher for other architectures.
There is also as yet no fully developed theory of error correction in this continuous-time setting, although quantum error correcting codes can be used to to provide robustness 
\cite{Lida2019},
and NMR and other quantum control techniques can improve the fidelity of the time evolution.
Even modest numbers of qubits can be used as quantum co-processors for bottleneck subroutines.
This is a natural way to exploit quantum computers in the early stages of development, through hybrid algorithms that use quantum computing only for the steps that are hardest for classical computers to do efficiently
\cite{Chan2016,Chan2017}.
More work is needed to develop theories of hybrid computational hardware, to enable the combinations to be fully exploited
\cite{Kend2011,Kend2013}.

%%%%%%%%%%%%%%%%%%%%%%%%%%%%%%%%%%%%%%%%%%%%%%%%%%%%%%%%%%%%%%%%%%%%%%%%%%%%%%
\subsection*{Acknowledgements}
This work was supported by the UKRI Engineering and Physical Science Council fellowship grant number EP/L022303/1.
Thanks to Jemma Bennett and Stephanie Foulds for interesting discussions and checking the manuscript.
A recording of the presentation of this work is available at \cite{Kend2020}.
%May as well get it into researchfish for something.

%%%%%%%%%%%%%%%%%%%%%%%%%%%%%%%%%%%%%%%%%%%%%%%%%%%%%%%%%%%%%%%%%%%%%%%%%%%%%%
\bibliographystyle{eptcs}
\bibliography{qwcomp}
\end{document}

%% file: Figure1.tex
%%%%%%%%%%%%%
%  FIG 1
%%%%%%%%%%%%%

\begin{minipage}[c]{0.45\linewidth}

\begin{tikzpicture}%[font=\large]

\draw[style=dashed] (0,0) -- (6,0);
\draw (0,0.5) node {$\scriptstyle{abstract}$};
\draw (0,-0.5) node {$\scriptstyle{physical}$};

\node[circle,draw] at (3,-1.5) {$e^-$};
\draw (4,1.25) node {$\psi: \ i \hbar \frac{\partial \psi}{\partial t} = H \psi$};

\end{tikzpicture}

(a)
%\textit{Spaces of abstract and physical objects (here, an electron and a wavefunction).}
\end{minipage}
\hfill
\begin{minipage}[c]{0.45\linewidth}

\begin{tikzpicture}%[font=\large]

\draw[style=dashed] (0,0) -- (6,0);
\draw (0,0.5) node {$\scriptstyle{abstract}$};
\draw (0,-0.5) node {$\scriptstyle{physical}$};

\node[circle,draw] at (3,-1.5) {$e^-$};
\draw (4,1.25) node {$\psi: \ i \hbar \frac{\partial \psi}{\partial t} = H \psi$};

\draw[->,double] (3,-1) -- (3,1);
\draw (3.5,0.25) node {$\mathcal{R}$};

\end{tikzpicture}

(b)
\end{minipage}

%% file: Figure2.tex
%%%%%%%%%%%%%
%  FIG 2
%%%%%%%%%%%%%
%Parallel evolution of theory and experiment.

\begin{minipage}[c]{0.45\linewidth}
\begin{tikzpicture}%[font=\large]

\draw[style=dashed] (0,0) -- (6,0);
\draw (0,0.5) node {$\scriptstyle{abstract}$};
\draw (0,-0.5) node {$\scriptstyle{physical}$};

%fig a
\node[draw] (p) at (1,-1) {$\mathbf{p}$};
\node[draw] (mp) at (1,2) {$m_{\mathbf{p}}$};
\draw[->,double] (p) -- (mp);
\draw (1.3,0.5) node {$\RT$};

\end{tikzpicture}

(a)
%\textit{Physical system $\p$ is represented abstractly by $m_\p$ using the modelling representation relation $\RT$ of theory $\theory$.}
\end{minipage}
\hfill
\begin{minipage}[c]{0.45\linewidth}
\begin{tikzpicture}%[font=\large]

\draw[style=dashed] (0,0) -- (6,0);
\draw (0,0.5) node {$\scriptstyle{abstract}$};
\draw (0,-0.5) node {$\scriptstyle{physical}$};

%fig a
\node[draw] (p) at (1,-1) {$\mathbf{p}$};
\node[draw] (mp) at (1,2) {$m_{\mathbf{p}}$};
\draw[->,double] (p) -- (mp);
\draw (1.3,0.5) node {$\RT$};

%fig b
\node[draw] (mprp) at (4.5,2) {$m^\prime_{\mathbf{p}}$};
\draw[-open triangle 45] (mp) -- (mprp);
\draw (2.8,2.5) node {$C_\theory(m_\mathbf{p})$};

\end{tikzpicture}

(b)
%\textit{Abstract dynamics $\see_\theory (m_\p)$ give the evolved abstract state $m^\prime_\p $.}
\end{minipage}

\begin{minipage}[c]{0.45\linewidth}
\begin{tikzpicture}%[font=\large]

\draw[style=dashed] (0,0) -- (6,0);
\draw (0,0.5) node {$\scriptstyle{abstract}$};
\draw (0,-0.5) node {$\scriptstyle{physical}$};

%fig a
\node[draw] (p) at (1,-1) {$\mathbf{p}$};
\node[draw] (mp) at (1,2) {$m_{\mathbf{p}}$};
\draw[->,double] (p) -- (mp);
\draw (1.3,0.5) node {$\mathcal{R}_\theory$};

%fig b
\node[draw] (mprp) at (4.5,2) {$m^\prime_{\mathbf{p}}$};
\draw[-open triangle 45] (mp) -- (mprp);
\draw (2.8,2.5) node {$C_\theory(m_\mathbf{p})$};

%fig c
\node[draw] (ppr) at (5,-1) {$\mathbf{p}^\prime$};
\draw[-open triangle 45] (p) -- (ppr);
\draw (3.25,-.75) node {$\mathbf{H}(\mathbf{p})$};

\end{tikzpicture}

(c)
%\textit{Physical dynamics $\mathbf{H}(\p)$ give the final physical state $\p^\prime$.}
\end{minipage}
\hfill
\begin{minipage}[c]{0.45\linewidth}
\begin{tikzpicture}%[font=\large]

\draw[style=dashed] (0,0) -- (6,0);
\draw (0,0.5) node {$\scriptstyle{abstract}$};
\draw (0,-0.5) node {$\scriptstyle{physical}$};

%fig a
\node[draw] (p) at (1,-1) {$\mathbf{p}$};
\node[draw] (mp) at (1,2) {$m_{\mathbf{p}}$};
\draw[->,double] (p) -- (mp);
\draw (1.3,0.5) node {$\RT$};

%fig b
\node[draw] (mprp) at (4.5,2) {$m^\prime_{\mathbf{p}}$};
\draw[-open triangle 45] (mp) -- (mprp);
\draw (2.8,2.5) node {$C_\theory(m_\mathbf{p})$};

%fig c
\node[draw] (ppr) at (5,-1) {$\mathbf{p}^\prime$};
\draw[-open triangle 45] (p) -- (ppr);
\draw (3.25,-.75) node {$\mathbf{H}(\mathbf{p})$};

%fig d
\node[draw] (mppr) at (5,1.2) {$m_{\mathbf{p}^\prime}$};
\draw[->,double] (ppr) -- (mppr);
\draw (5.3,0.3) node {$\RT$};

\draw[<->] (mppr) to[bend right=45] (mprp);
\draw (5.3,1.8) node {$\varepsilon$};

\end{tikzpicture}

(d)
%\textit{$\RT$ is used again to represent $\p^\prime$ as $m_{\p^\prime}$.}
\end{minipage}

%% file: Figure3.tex
%%%%%%%%%%%%%
%  FIG 3 - prediction
%%%%%%%%%%%%%

\begin{minipage}[c]{0.45\linewidth}

\begin{tikzpicture}%[font=\large]
   
\draw[style=dashed] (0,0) -- (6,0);
\draw (0,0.5) node {$\scriptstyle{abstract}$};
\draw (0,-0.4) node {$\scriptstyle{physical}$};

\node[draw] (p) at (1,-1) {$\p$};
\node[draw] (mp) at (1,2) {$m_\p$};
\draw[->,double] (p) -- (mp);
\draw (1.5,0.5) node {$\RT$};

\node[draw] (mprp) at (4.5,2) {$m^\prime_\p$};

\draw[-open triangle 45] (mp) -- (mprp);

\end{tikzpicture}

(a)
%\textit{The `predict process': abstract theory is used to predict physical evolution.}

\end{minipage}
\hfill
\begin{minipage}[c]{0.45\linewidth}

\begin{tikzpicture}%[font=\large]
   
\draw[style=dashed] (0,0) -- (6,0);
\draw (0,0.5) node {$\scriptstyle{abstract}$};
\draw (0,-0.4) node {$\scriptstyle{physical}$};

%\node[draw] (p) at (1,-1) {$\p$};
\node[draw] (mp) at (1,2) {$m_\p$};
%\draw[->,double] (p) -- (mp);
%\draw (1.5,0.5) node {$\RT$};

\node[draw] (mprp) at (5,2) {$m^\prime_\p\approx m_{\p^\prime}$};

\draw[-open triangle 45] (mp) -- (mprp);

\node[draw] (ppr) at (5,-1) {$\p^\prime$};

\draw[->,double] (mprp) -- (ppr);
\draw (5.5,0.5) node {$\IRT$};

\end{tikzpicture}

(b)
%\textit{The `instantiation process': using an instantiation representation relation $\IRT$, a physical object is found corresponding to the predicted evolution.}

\end{minipage}

%% file: Figure4.tex
%%%%%%%%%%%%%
%  FIG 4 - engineering
%%%%%%%%%%%%%

\begin{tikzpicture}%[font=\large]

\draw[style=dashed] (0,0) -- (6,0);
\draw (0,0.5) node {$\scriptstyle{abstract}$};
\draw (0,-0.4) node {$\scriptstyle{physical}$};

%fig a
\node[draw] (p) at (1,-1) {$\mathbf{p}$};
\draw (0.2,-0.9) node {\small raw};
\draw (-0.1,-1.2) node {\small material};
\node[draw] (mp) at (1,2) {$m_{\mathbf{p}}$};
\draw[->,double] (mp) -- (p);
\draw (1.3,0.5) node {$\IRT$};

%fig b
\node[draw] (mprp) at (4.5,2) {$m^\prime_{\mathbf{p}}$};
\draw[-open triangle 45] (mp) -- (mprp);
\draw (2.8,2.5) node {$C_\theory(m_\mathbf{p})$};

%fig c
\node[draw] (ppr) at (5,-1) {$\mathbf{p}^\prime$};
\draw (6.2,-0.9) node {\small finished};
\draw (6.2,-1.2) node {\small product};
\draw[-open triangle 45] (p) -- (ppr);
\draw (3.25,-.75) node {$\mathbf{H}(\mathbf{p})$};

%fig d
\node[draw] (mppr) at (5,1.2) {$m_{\mathbf{p}^\prime}$};
\draw[->,double] (ppr) -- (mppr);
\draw (5.3,0.3) node {$\RT$};

\draw[<->] (mppr) to[bend right=45] (mprp);
\draw (5.3,1.8) node {$\varepsilon$};

%Making an artifact $p^{\prime}$ from a design specification $C_{\theory}(m_{\p})$ and raw material $\p$.

\end{tikzpicture}

%% file: Figure6.tex
%%%%%%%%%%%%%
%  FIG 5
%%%%%%%%%%%%%

%\begin{minipage}[c]{1.0\linewidth}
\begin{tikzpicture}%[font=\large]
   
\draw[style=dashed] (1,0) -- (8,0);
\draw (1,0.5) node {$\scriptstyle{abstract}$};
\draw (1,-0.5) node {$\scriptstyle{physical}$};

\node[draw] (p) at (3,-1) {$\p$};
\node[draw] (mp) at (3,2) {$m_\p$};
\draw[->,double] (mp) -- (p);
\draw (3.8,0.5) node {$\IRT$};

\node[draw] (mprp) at (7,2) {$ m_{\p^\prime} \approx m^\prime_\p$};

\node[draw] (ppr) at (7,-1) {$\p^\prime$};

\draw[-open triangle 45] (p) -- (ppr);

\draw (5.25,-.75) node {$\mathbf{H}(\p)$};

\draw[->,double] (ppr) -- (mprp);
\draw (7.75,0.5) node {$\RT$};

\node[draw] (ms) at (1,2.5) {$m_\mathbf{s}$};
 
\draw[->] (ms) to[bend left=30] (mp); 
\draw (2.25,2.95) node {$\Delta$}; 
   
\node[draw] (mspr) at (9,3) {$m^\prime_\mathbf{s}$};
 
\draw[->] (mprp) to[bend left=30] (mspr); 
\draw (7.75,3.25) node {$\tilde{\Delta}$};

\end{tikzpicture}

%\end{minipage}

%% file: Figure8.tex
%%%%%%%%%%%%%
%  FIG 8
%%%%%%%%%%%%%

\begin{minipage}{0.45\linewidth}
\begin{tikzpicture}%[font=\large]  
    \draw[style=dashed] (0,0) -- (6,0);
    \draw (0,0.5) node {$\scriptstyle{abstract}$};
    \draw (0,-0.5) node {$\scriptstyle{physical}$};

   \node[draw] (p) at (1,-1.5) {$\p$};
   
   \node[draw] (mp) at (1,1.5) {$m_\p$};
   \node[draw] (mprp) at (5,1.5) {$ m^\prime_\p$};

   \draw[->,double] (p) -- (mp);
   \draw[-open triangle 45] (mp) -- (mprp);
   
   \draw (1.5,-0.5) node {$\mathcal{R}_{\theory(\p)}$};
   \draw (3,2.0) node {$C_\theory(\p)$};
    
\end{tikzpicture}

(a)
%\textit{Predict process using $C(\p)$ to find the abstract prediction for the evolution of system $\p$}
\end{minipage}
\hfill
\begin{minipage}[c]{0.47\linewidth}
\begin{tikzpicture}%[font=\large]
    \draw[style=dashed] (0,0) -- (6,0);
    \draw (0,0.5) node {$\scriptstyle{abstract}$};
    \draw (0,-0.5) node {$\scriptstyle{physical}$};
  
   \node[draw] (p) at (1,-1.5) {$\p$};
   
   \node[draw] (ns) at (2.5,1) {$n_\mathbf{s}$};
   \node[draw] (mp) at (1,1.5) {$m_\p$};
   \node[draw] (nspr) at (4.5,1) {$n^\prime_\mathbf{s}$};
   \node[draw] (mppr) at (6,1.5) {$ m^\prime_\p$};
   
   \draw[-open triangle 45] (ns) -- (nspr);
   \draw[->,double] (p) -- (mp);
   
   \draw[->] (mp) to[bend left=30] (ns);
   \draw[->] (nspr) to[bend left=30] (mppr);
   
   \draw (3.5,1.5) node {$C_\theory (\mathbf{s})$};
   \draw (1.5,-0.5) node {$\mathcal{R}_{\theory(\p)}$};
   
   \draw (5.1,1.7) node {$\tilde{\Delta}$};
   \draw (1.9,1.7) node {$\Delta$};
    
 \end{tikzpicture}
 
(b)
%\textit{Encoding $C(\p)$ in the simulator model dynamics, $C(\mathbf{s})$ using $E$ and $D$}
\end{minipage}

\begin{minipage}[c]{0.45\linewidth}
\begin{tikzpicture}%[font=\large] 
	\draw[style=dashed] (0,0) -- (6,0);
    \draw (0,0.5) node {$\scriptstyle{abstract}$};
    \draw (0,-0.5) node {$\scriptstyle{physical}$};
  
   \node[draw] (s) at (2.5,-1) {$\mathbf{s}$};
   \node[draw] (p) at (1,-1.5) {$\p$};
   \node[draw] (spr) at (4.5,-1) {$\mathbf{s}^\prime$};
   
   \node[draw] (ns) at (2.5,1) {$n_\mathbf{s}$};
   \node[draw] (mp) at (1,1.5) {$m_\p$};
   \node[draw] (nspr) at (4.5,1) {$n_{\mathbf{s}^\prime}\approx n^\prime_\mathbf{s}$};
   \node[draw] (mppr) at (6,1.5) {$ m^\prime_\p$};
   
   \draw[-open triangle 45] (s) -- (spr);
   \draw[->,double] (ns) -- (s);
   \draw[->,double] (p) -- (mp);
   \draw[->,double] (spr) -- (nspr);
   
   \draw[->] (mp) to[bend left=30] (ns);
   \draw[->] (nspr) to[bend left=30] (mppr);
   
   \draw (1.5,-0.5) node {$\mathcal{R}_{\theory(\p)}$};
   \draw (3,-0.4) node {$\tilde{\mathcal{R}}_{\theory(\mathbf{s})}$};
   \draw (5,-0.4) node {$\mathcal{R}_{\theory(\mathbf{s})}$};
   
   \draw (5.1,1.7) node {$\tilde{\Delta}$};
   \draw (1.9,1.7) node {$\Delta$};
    
\end{tikzpicture}

(c)
%\textit{Adding the compute process: the physical simulator $\mathbf{s}$ now determines the abstract evolution $C(\mathbf{s})$ and via encode/decode $E/D$ calculates $C(\p)$}
\end{minipage}
\hfill
\begin{minipage}[c]{0.47\linewidth}
\begin{tikzpicture}%[font=\large]
    \draw[style=dashed] (0,0) -- (6,0);
    \draw (0,0.5) node {$\scriptstyle{abstract}$};
    \draw (0,-0.5) node {$\scriptstyle{physical}$};
  
   \node[draw] (p) at (1,-1.5) {$\p=\mathbf{s}$};
   \node[draw] (ppr) at (5.5,-1.5) {$\p^\prime=\mathbf{s}^\prime$};
   
   \node[draw] (mp) at (1,1.5) {$m_\p = n_\mathbf{s}$};
   \node[draw] (mppr) at (5.5,1.5) {$ m_{\p^\prime} \approx m^\prime_\p = n^\prime_\mathbf{s}$};

   \draw[<->,double] (p) -- (mp);
   \draw[<->,double] (ppr) -- (mppr);
   \draw[-open triangle 45] (p) -- (ppr);
   
   \draw (3,-0.4) node {$\{\mathcal{R}_{\theory(\p)}$ and $\mathcal{\widetilde{R}}_{\theory(\p)}\}$};
   \draw (3.25, -1.0) node {$=\{\mathcal{R}_{\theory(\mathbf{s})}$ and $\mathcal{\widetilde{R}}_{\theory(\mathbf{s})}\}$};
%   \draw (6.8,-0.3) node {$\{\mathcal{R}_{\theory(\p)},\mathcal{\widetilde{R}}_{\theory(\p)}\}$};
%   \draw (6.5, -1.2) node {$=\{\mathcal{R}_{\theory(\mathbf{s})},\mathcal{\widetilde{R}}_{\theory(\mathbf{s})}\}$};

	\end{tikzpicture}

(d)
% self-sim - A system simulating itself, with encode and decode as identities.
\end{minipage}

%% file: hypercubefig.tex
\begin{tikzpicture}%[font=\large]

    %%% three qubits %%%
	\draw[<-,very thick] (11.5,1) -- (11.6,0.5);
    \draw[<-,very thick] (12.5,1) -- (12.6,0.5);
    \draw[<-,very thick] (12.0,1) -- (12.1,1.5);

	\draw[red,dashed] (12.55,0.75) -- (11.55,0.75);
    \draw[red,dashed] (11.55,0.75) -- (12.05,1.25);
    \draw[red,dashed] (12.05,1.25) -- (12.55,0.75);

	%%% cube %%%
	\draw (10.5,1.0) node {$N=8$};
	\draw (10.5,0.5) node {$n=3$};

	\draw (11,3) rectangle (14,6);
	\fill[black] (11,3) circle (1ex);
	\fill[black] (11,6) circle (1ex);
	\fill[black] (14,3) circle (1ex);
	\fill[black] (14,6) circle (1ex);
	
	\draw (10,2) -- (11,3);
	\draw (10,5) -- (11,6);
	\draw (13,2) -- (14,3);
	\draw (13,5) -- (14,6);

	\draw (11.5,3.4) node {100};
	\draw (11.5,6.4) node {110};
	\draw (14.5,3.4) node {101};
	\draw (14.5,6.4) node {111};

	\draw (10,2) rectangle (13,5);
	\fill[black] (10,2) circle (1ex);
	\fill[black] (10,5) circle (1ex);
	\fill[black] (13,2) circle (1ex);
	\fill[black] (13,5) circle (1ex);

	\draw (9.7,1.6) node {000};
	\draw (9.7,5.3) node {010};
	\draw (13.3,1.6) node {001};
	\draw (13.5,5.0) node {011};

    %%% two qubits %%%
	\draw[<-,very thick] (6.5,1) -- (6.6,0.5);
    \draw[->,very thick] (7.5,1) -- (7.6,0.5);

    \draw[red,dashed] (6.55,0.75) -- (7.55,0.75);

	%%% square %%%
	\draw (5.5,1.0) node {$N=4$};
	\draw (5.5,0.5) node {$n=2$};

	\draw (5,2) rectangle (8,5);
	\fill[black] (5,2) circle (1ex);
	\fill[black] (5,5) circle (1ex);
	\fill[black] (8,2) circle (1ex);
	\fill[black] (8,5) circle (1ex);

	\draw (4.5,1.7) node {00};
	\draw (4.5,5.2) node {01};
	\draw (8.5,1.7) node {10};
	\draw (8.5,5.2) node {11};

    %%% one qubit %%%
	\draw[<-,very thick] (1.9,1.0) -- (2,0.5);

	%%% line %%%
	\draw (0.5,1.0) node {$N=2$};
	\draw (0.5,0.5) node {$n=1$};

	\draw (0,2) -- (3,2);

	\fill[black] (0,2) circle (1ex);
	\fill[black] (3,2) circle (1ex);

	\draw (-0.3,1.6) node {0};
	\draw (3.3,1.6) node {1};

\end{tikzpicture}

%% file: qwcomp.bbl
\begin{thebibliography}{10}
\providecommand{\bibitemdeclare}[2]{}
\providecommand{\surnamestart}{}
\providecommand{\surnameend}{}
\providecommand{\urlprefix}{Available at }
\providecommand{\url}[1]{\texttt{#1}}
\providecommand{\href}[2]{\texttt{#2}}
\providecommand{\urlalt}[2]{\href{#1}{#2}}
\providecommand{\doi}[1]{doi:\urlalt{http://dx.doi.org/#1}{#1}}
\providecommand{\bibinfo}[2]{#2}

\bibitemdeclare{article}{Aaro2010}
\bibitem{Aaro2010}
\bibinfo{author}{Scott \surnamestart Aaronson\surnameend} \&
  \bibinfo{author}{Alex \surnamestart Arkhipov\surnameend}
  (\bibinfo{year}{2013}): \emph{\bibinfo{title}{The Computational Complexity of
  Linear Optics}}.
\newblock {\sl \bibinfo{journal}{Theory of Computing}}
  \bibinfo{volume}{9}(\bibinfo{number}{4}), pp. \bibinfo{pages}{143--252},
  \doi{10.4086/toc.2013.v009a004}.

\bibitemdeclare{article}{Ahar2004}
\bibitem{Ahar2004}
\bibinfo{author}{Dorit \surnamestart Aharonov\surnameend}, \bibinfo{author}{Wim
  \surnamestart van Dam\surnameend}, \bibinfo{author}{Julia \surnamestart
  Kempe\surnameend}, \bibinfo{author}{Zeph \surnamestart Landau\surnameend},
  \bibinfo{author}{Seth \surnamestart Lloyd\surnameend} \&
  \bibinfo{author}{Oded \surnamestart Regev\surnameend} (\bibinfo{year}{2007}):
  \emph{\bibinfo{title}{Adiabatic Quantum Computation is Equivalent to Standard
  Quantum Computation}}.
\newblock {\sl \bibinfo{journal}{SIAM Journal on Computing}}
  \bibinfo{volume}{37}, pp. \bibinfo{pages}{166--194},
  \doi{10.1137/S0097539705447323}.

\bibitemdeclare{article}{Benn1997}
\bibitem{Benn1997}
\bibinfo{author}{C.~\surnamestart Bennett\surnameend},
  \bibinfo{author}{E.~\surnamestart Bernstein\surnameend},
  \bibinfo{author}{G.~\surnamestart Brassard\surnameend} \&
  \bibinfo{author}{U.~\surnamestart Vazirani\surnameend}
  (\bibinfo{year}{1997}): \emph{\bibinfo{title}{Strengths and {W}eaknesses of
  {Q}uantum {C}omputing}}.
\newblock {\sl \bibinfo{journal}{SIAM Journal on Computing}}
  \bibinfo{volume}{26}(\bibinfo{number}{5}), pp. \bibinfo{pages}{1510--1523},
  \doi{10.1137/S0097539796300933}.

\bibitemdeclare{article}{Blum2002}
\bibitem{Blum2002}
\bibinfo{author}{Robin \surnamestart Blume-Kohout\surnameend},
  \bibinfo{author}{Carlton~M. \surnamestart Caves\surnameend} \&
  \bibinfo{author}{Ivan~H. \surnamestart Deutsch\surnameend}
  (\bibinfo{year}{2002}): \emph{\bibinfo{title}{Climbing Mount Scalable:
  Physical Resource Requirements for a Scalable Quantum Computer}}.
\newblock {\sl \bibinfo{journal}{Found.~Phys.}}
  \bibinfo{volume}{32}(\bibinfo{number}{11}), pp. \bibinfo{pages}{1641--1670},
  \doi{10.1023/A:1021471621587}.

\bibitemdeclare{article}{Boug2016}
\bibitem{Boug2016}
\bibinfo{author}{Hamza \surnamestart Bougroura\surnameend},
  \bibinfo{author}{Habib \surnamestart Aissaoui\surnameend},
  \bibinfo{author}{Nicholas \surnamestart Chancellor\surnameend} \&
  \bibinfo{author}{Viv \surnamestart Kendon\surnameend} (\bibinfo{year}{2016}):
  \emph{\bibinfo{title}{Quantum-walk transport properties on graphene
  structures}}.
\newblock {\sl \bibinfo{journal}{Phys. Rev. A}} \bibinfo{volume}{94}, p.
  \bibinfo{pages}{062331}, \doi{10.1103/PhysRevA.94.062331}.

\bibitemdeclare{article}{Broo2010}
\bibitem{Broo2010}
\bibinfo{author}{M.~A. \surnamestart Broome\surnameend},
  \bibinfo{author}{A.~\surnamestart Fedrizzi\surnameend},
  \bibinfo{author}{B.~P. \surnamestart Lanyon\surnameend},
  \bibinfo{author}{I.~\surnamestart Kassal\surnameend},
  \bibinfo{author}{A.~\surnamestart Aspuru-Guzik\surnameend} \&
  \bibinfo{author}{A.~G. \surnamestart White\surnameend}
  (\bibinfo{year}{2010}): \emph{\bibinfo{title}{Discrete single-photon quantum
  walks with tunable decoherence}}.
\newblock {\sl \bibinfo{journal}{Phys.~Rev.~Lett.}} \bibinfo{volume}{104}, p.
  \bibinfo{pages}{153602}, \doi{10.1103/PhysRevLett.104.153602}.

\bibitemdeclare{article}{Brow2010}
\bibitem{Brow2010}
\bibinfo{author}{K.~L. \surnamestart Brown\surnameend}, \bibinfo{author}{W.~J.
  \surnamestart Munro\surnameend} \& \bibinfo{author}{V.~M. \surnamestart
  Kendon\surnameend} (\bibinfo{year}{2010}): \emph{\bibinfo{title}{Using
  Quantum Computers for Quantum Simulation}}.
\newblock {\sl \bibinfo{journal}{Entropy}}
  \bibinfo{volume}{12}(\bibinfo{number}{11}), pp. \bibinfo{pages}{2268--2307},
  \doi{10.3390/e12112268}.

\bibitemdeclare{article}{Call2019}
\bibitem{Call2019}
\bibinfo{author}{A~\surnamestart Callison\surnameend},
  \bibinfo{author}{N~\surnamestart Chancellor\surnameend},
  \bibinfo{author}{F~\surnamestart Mintert\surnameend} \&
  \bibinfo{author}{V~\surnamestart Kendon\surnameend} (\bibinfo{year}{2019}):
  \emph{\bibinfo{title}{Finding spin glass ground states using quantum walks}}.
\newblock {\sl \bibinfo{journal}{New J.~Phys.}} \bibinfo{volume}{21}, p.
  \bibinfo{pages}{123022}, \doi{10.1088/1367-2630/ab5ca2}.

\bibitemdeclare{misc}{Chan2017}
\bibitem{Chan2017}
\bibinfo{author}{N.~\surnamestart Chancellor\surnameend}
  (\bibinfo{year}{2017}): \emph{\bibinfo{title}{Modernizing Quantum Annealing
  II: Genetic algorithms with the Inference Primitive Formalism}}.
\newblock \urlprefix\url{https://arxiv.org/abs/1609.05875}.
\newblock \bibinfo{note}{ArXiv:1609.05875}.

\bibitemdeclare{article}{Chan2016}
\bibitem{Chan2016}
\bibinfo{author}{N.~\surnamestart Chancellor\surnameend}
  (\bibinfo{year}{2017}): \emph{\bibinfo{title}{Modernizing Quantum Annealing
  using Local Searches}}.
\newblock {\sl \bibinfo{journal}{New J.~Phys.}}
  \bibinfo{volume}{19}(\bibinfo{number}{2}), p. \bibinfo{pages}{023024},
  \doi{10.1088/1367-2630/aa59c4}.

\bibitemdeclare{article}{Chil2009}
\bibitem{Chil2009}
\bibinfo{author}{Andrew~M. \surnamestart Childs\surnameend}
  (\bibinfo{year}{2009}): \emph{\bibinfo{title}{Universal computation by
  quantum walk}}.
\newblock {\sl \bibinfo{journal}{Phys.~Rev.~Lett.}} \bibinfo{volume}{102}, p.
  \bibinfo{pages}{180501}, \doi{10.1103/PhysRevLett.102.180501}.

\bibitemdeclare{inproceedings}{Chil2002}
\bibitem{Chil2002}
\bibinfo{author}{Andrew~M. \surnamestart Childs\surnameend},
  \bibinfo{author}{Richard \surnamestart Cleve\surnameend},
  \bibinfo{author}{Enrico \surnamestart Deotto\surnameend},
  \bibinfo{author}{Edward \surnamestart Farhi\surnameend}, \bibinfo{author}{Sam
  \surnamestart Gutmann\surnameend} \& \bibinfo{author}{Daniel~A. \surnamestart
  Spielman\surnameend} (\bibinfo{year}{2003}):
  \emph{\bibinfo{title}{Exponential algorithmic speedup by a quantum walk}}.
\newblock In: {\sl \bibinfo{booktitle}{Proc.~35th Annual ACM Symposium on
  Theory of Computing (STOC 2003)}}, \bibinfo{publisher}{Assoc.~for
  Comp.~Machinery, New York}, pp. \bibinfo{pages}{59--68},
  \doi{10.1145/780542.780552}.

\bibitemdeclare{article}{Chil2001}
\bibitem{Chil2001}
\bibinfo{author}{Andrew~M. \surnamestart Childs\surnameend},
  \bibinfo{author}{Edward \surnamestart Farhi\surnameend} \&
  \bibinfo{author}{John \surnamestart Preskill\surnameend}
  (\bibinfo{year}{2001}): \emph{\bibinfo{title}{{Robustness of adiabatic
  quantum computation}}}.
\newblock {\sl \bibinfo{journal}{Physical Review A}}
  \bibinfo{volume}{65}(\bibinfo{number}{1}), p. \bibinfo{pages}{012322},
  \doi{10.1103/PhysRevA.65.012322}.

\bibitemdeclare{article}{Chil2004}
\bibitem{Chil2004}
\bibinfo{author}{Andrew~M. \surnamestart Childs\surnameend} \&
  \bibinfo{author}{Jeffrey \surnamestart Goldstone\surnameend}
  (\bibinfo{year}{2004}): \emph{\bibinfo{title}{{Spatial search by quantum
  walk}}}.
\newblock {\sl \bibinfo{journal}{Physical Review A}}
  \bibinfo{volume}{70}(\bibinfo{number}{2}), p. \bibinfo{pages}{022314},
  \doi{10.1103/PhysRevA.70.022314}.

\bibitemdeclare{article}{Chil2013}
\bibitem{Chil2013}
\bibinfo{author}{Andrew~M. \surnamestart Childs\surnameend},
  \bibinfo{author}{David \surnamestart Gosset\surnameend} \&
  \bibinfo{author}{Zak \surnamestart Webb\surnameend} (\bibinfo{year}{2013}):
  \emph{\bibinfo{title}{Universal computation by multi-particle quantum walk}}.
\newblock {\sl \bibinfo{journal}{Science}} \bibinfo{volume}{339}, pp.
  \bibinfo{pages}{791--794}, \doi{10.1126/science.1229957}.

\bibitemdeclare{misc}{Choi2010}
\bibitem{Choi2010}
\bibinfo{author}{Vicky \surnamestart Choi\surnameend} (\bibinfo{year}{2010}):
  \emph{\bibinfo{title}{Adiabatic quantum algorithms for the {NP}-complete
  {M}aximum-{W}eight {I}ndependent set, {E}xact {C}over and 3{SAT} problems}}.
\newblock \urlprefix\url{https://arxiv.org/abs/1004.2226}.
\newblock \bibinfo{note}{ArXiv:1004.2226}.

\bibitemdeclare{article}{Dodd2019}
\bibitem{Dodd2019}
\bibinfo{author}{A.~Ben \surnamestart Dodds\surnameend}, \bibinfo{author}{Viv
  \surnamestart Kendon\surnameend}, \bibinfo{author}{Charles~S. \surnamestart
  Adams\surnameend} \& \bibinfo{author}{Nicholas \surnamestart
  Chancellor\surnameend} (\bibinfo{year}{2019}):
  \emph{\bibinfo{title}{Practical designs for permutation-symmetric problem
  Hamiltonians on hypercubes}}.
\newblock {\sl \bibinfo{journal}{Phys. Rev. A}} \bibinfo{volume}{100}, p.
  \bibinfo{pages}{032320}, \doi{10.1103/PhysRevA.100.032320}.

\bibitemdeclare{article}{Eker1998}
\bibitem{Eker1998}
\bibinfo{author}{A~\surnamestart Ekert\surnameend} \&
  \bibinfo{author}{J~\surnamestart Jozsa\surnameend} (\bibinfo{year}{1998}):
  \emph{\bibinfo{title}{Quantum algorithms: entanglement–enhanced information
  processing}}.
\newblock {\sl \bibinfo{journal}{Phil.~Trans.~Royal Soc.~A}}
  \bibinfo{volume}{356}, pp. \bibinfo{pages}{1769--82},
  \doi{10.1098/rsta.1998.0248}.

\bibitemdeclare{misc}{Farh2000}
\bibitem{Farh2000}
\bibinfo{author}{E.~\surnamestart Farhi\surnameend},
  \bibinfo{author}{J.~\surnamestart Goldstone\surnameend},
  \bibinfo{author}{S.~\surnamestart Gutmann\surnameend} \&
  \bibinfo{author}{M.~\surnamestart Sipser\surnameend} (\bibinfo{year}{2000}):
  \emph{\bibinfo{title}{Quantum Computation by Adiabatic Evolution}}.
\newblock \urlprefix\url{http://arxiv.org/quant-ph/abs/0001106}.
\newblock \bibinfo{note}{ArXiv:quant-ph/0001106}.

\bibitemdeclare{article}{Farh1998}
\bibitem{Farh1998}
\bibinfo{author}{E~\surnamestart Farhi\surnameend} \&
  \bibinfo{author}{S~\surnamestart Gutmann\surnameend} (\bibinfo{year}{1998}):
  \emph{\bibinfo{title}{Quantum computation and decison trees}}.
\newblock {\sl \bibinfo{journal}{Phys.~Rev.~A}} \bibinfo{volume}{58}, pp.
  \bibinfo{pages}{915--928}, \doi{10.1103/PhysRevA.58.915}.

\bibitemdeclare{article}{Fini1994}
\bibitem{Fini1994}
\bibinfo{author}{A.~B. \surnamestart Finilla\surnameend},
  \bibinfo{author}{M.~A. \surnamestart Gomez\surnameend},
  \bibinfo{author}{C.~\surnamestart Sebenik\surnameend} \&
  \bibinfo{author}{J.~D. \surnamestart Doll\surnameend} (\bibinfo{year}{1994}):
  \emph{\bibinfo{title}{Quantum annealing: A new method for minimizing
  multidimensional functions}}.
\newblock {\sl \bibinfo{journal}{Chem. Phys. Lett.}} \bibinfo{volume}{219}, p.
  \bibinfo{pages}{343}, \doi{10.1016/0009-2614(94)00117-0}.

\bibitemdeclare{misc}{Hard2001}
\bibitem{Hard2001}
\bibinfo{author}{Lucien \surnamestart Hardy\surnameend} (\bibinfo{year}{2001}):
  \emph{\bibinfo{title}{Quantum theory from five reasonable axioms}}.
\newblock \urlprefix\url{http://arxiv.org/quant-ph/abs/0101012}.
\newblock \bibinfo{note}{ArXiv:quant-ph/0101012}.

\bibitemdeclare{article}{Hart1984}
\bibitem{Hart1984}
\bibinfo{author}{A.~\surnamestart Hartwig\surnameend},
  \bibinfo{author}{F.~\surnamestart Daske\surnameend} \&
  \bibinfo{author}{S.~\surnamestart Kobe\surnameend} (\bibinfo{year}{1984}):
  \emph{\bibinfo{title}{A recursive branch-and-bound algorithm for the exact
  ground state of Ising spin-glass models}}.
\newblock {\sl \bibinfo{journal}{Computer Physics Communications}}
  \bibinfo{volume}{32}(\bibinfo{number}{2}), pp. \bibinfo{pages}{133 -- 138},
  \doi{10.1016/0010-4655(84)90066-3}.

\bibitemdeclare{article}{Hors2014}
\bibitem{Hors2014}
\bibinfo{author}{C.~\surnamestart Horsman\surnameend},
  \bibinfo{author}{S.~\surnamestart Stepney\surnameend}, \bibinfo{author}{R.~C.
  \surnamestart Wagner\surnameend} \& \bibinfo{author}{V.~\surnamestart
  Kendon\surnameend} (\bibinfo{year}{2014}): \emph{\bibinfo{title}{When does a
  Physical System Compute?}}
\newblock {\sl \bibinfo{journal}{Proc.~Roy.~Soc.~A}}
  \bibinfo{volume}{470}(\bibinfo{number}{2169}), p. \bibinfo{pages}{20140182},
  \doi{10.1098/rspa.2014.0182}.

\bibitemdeclare{inproceedings}{Hors2017}
\bibitem{Hors2017}
\bibinfo{author}{D~\surnamestart Horsman\surnameend},
  \bibinfo{author}{V~\surnamestart Kendon\surnameend},
  \bibinfo{author}{S~\surnamestart Stepney\surnameend} \&
  \bibinfo{author}{P~\surnamestart Young\surnameend} (\bibinfo{year}{2017}):
  \emph{\bibinfo{title}{Abstraction and representation in living organisms:
  when does a biological system compute?}}
\newblock In \bibinfo{editor}{Giovagnoli~R \surnamestart
  Dodig-Crnkovic~G\surnameend}, editor: {\sl \bibinfo{booktitle}{Representation
  and Reality in Humans, Other Living Organisms and Intelligent Machines}},
  {\sl \bibinfo{series}{Studies in Applied Philosophy, Epistemology and
  Rational Ethics}}~\bibinfo{volume}{28}, \bibinfo{publisher}{Springer}, pp.
  \bibinfo{pages}{91--116}, \doi{10.1007/978-3-319-43784-2_6}.

\bibitemdeclare{article}{Hugh1997}
\bibitem{Hugh1997}
\bibinfo{author}{Richard I.~G. \surnamestart Hughes\surnameend}
  (\bibinfo{year}{1997}): \emph{\bibinfo{title}{Models and representation}}.
\newblock {\sl \bibinfo{journal}{Philosophy of science}} \bibinfo{volume}{64},
  pp. \bibinfo{pages}{S325--S336}, \doi{10.1086/392611}.

\bibitemdeclare{article}{Kado1998}
\bibitem{Kado1998}
\bibinfo{author}{T.~\surnamestart Kadowaki\surnameend} \&
  \bibinfo{author}{H.~\surnamestart Nishimori\surnameend}
  (\bibinfo{year}{1998}): \emph{\bibinfo{title}{Quantum annealing in the
  transverse Ising model}}.
\newblock {\sl \bibinfo{journal}{Phys. Rev. E}} \bibinfo{volume}{58}, p.
  \bibinfo{pages}{5355}, \doi{10.1103/PhysRevE.58.5355}.

\bibitemdeclare{article}{Kars2009}
\bibitem{Kars2009}
\bibinfo{author}{Michal \surnamestart Karski\surnameend},
  \bibinfo{author}{Leonid \surnamestart Forster\surnameend},
  \bibinfo{author}{Jai-Min \surnamestart Choi\surnameend},
  \bibinfo{author}{Andreas \surnamestart Steffen\surnameend},
  \bibinfo{author}{Wolfgang \surnamestart Alt\surnameend},
  \bibinfo{author}{Dieter \surnamestart Meschede\surnameend} \&
  \bibinfo{author}{Artur \surnamestart Widera\surnameend}
  (\bibinfo{year}{2009}): \emph{\bibinfo{title}{{Quantum Walk in Position Space
  with Single Optically Trapped Atoms}}}.
\newblock {\sl \bibinfo{journal}{Science}}
  \bibinfo{volume}{325}(\bibinfo{number}{5937}), pp. \bibinfo{pages}{174--177},
  \doi{10.1126/science.1174436}.

\bibitemdeclare{inproceedings}{Kemp2004}
\bibitem{Kemp2004}
\bibinfo{author}{\surnamestart Kempe\surnameend},
  \bibinfo{author}{\surnamestart Kitaev\surnameend} \&
  \bibinfo{author}{\surnamestart Regev\surnameend} (\bibinfo{year}{2004}):
  \emph{\bibinfo{title}{The Complexity of the Local Hamiltonian Problem}}.
\newblock In \bibinfo{editor}{K.~\surnamestart Lodaya\surnameend} \&
  \bibinfo{editor}{M.~\surnamestart Mahajan\surnameend}, editors: {\sl
  \bibinfo{booktitle}{Proc.~24th FSTTCS}}, {\sl \bibinfo{series}{LNCS}}
  \bibinfo{volume}{3328}, \bibinfo{publisher}{Springer}, pp.
  \bibinfo{pages}{372--383}, \doi{10.1007/978-3-540-30538-5_31}.

\bibitemdeclare{inproceedings}{Kend2011}
\bibitem{Kend2011}
\bibinfo{author}{V.~\surnamestart Kendon\surnameend},
  \bibinfo{author}{A.~\surnamestart Sebald\surnameend},
  \bibinfo{author}{S.~\surnamestart Stepney\surnameend},
  \bibinfo{author}{M.~\surnamestart Bechmann\surnameend},
  \bibinfo{author}{P.~\surnamestart Hines\surnameend} \& \bibinfo{author}{R.~C.
  \surnamestart Wagner\surnameend} (\bibinfo{year}{2011}):
  \emph{\bibinfo{title}{Heterotic computing}}.
\newblock In \bibinfo{editor}{C.S. \surnamestart Calude\surnameend},
  \bibinfo{editor}{J.~\surnamestart Kari\surnameend},
  \bibinfo{editor}{I.~\surnamestart Petre\surnameend} \&
  \bibinfo{editor}{G.~\surnamestart Rozenberg\surnameend}, editors: {\sl
  \bibinfo{booktitle}{Unconventional Computation, LNCS}},
  \bibinfo{volume}{6714}, \bibinfo{publisher}{Springer},
  \bibinfo{address}{Berlin, Heidelberg}, pp. \bibinfo{pages}{113--124},
  \doi{10.1007/978-3-642-21341-0_16}.

\bibitemdeclare{proceedings}{Kend2013}
\bibitem{Kend2013}
\bibinfo{editor}{V~\surnamestart Kendon\surnameend},
  \bibinfo{editor}{A~\surnamestart Siebald\surnameend} \&
  \bibinfo{editor}{S~\surnamestart Stepney\surnameend}, editors
  (\bibinfo{year}{2015}): \emph{\bibinfo{title}{Heterotic computing: exploiting
  hybrid computational devices}}. {\sl \bibinfo{series}{Phil.~Trans.~Royal
  Soc.~A}} \bibinfo{volume}{373}, \bibinfo{publisher}{Royal Society},
  \bibinfo{address}{London, UK}, \doi{10.1098/rsta.2015.0091}.

\bibitemdeclare{misc}{Kend2020}
\bibitem{Kend2020}
\bibinfo{author}{Viv \surnamestart Kendon\surnameend} (\bibinfo{year}{2020}):
  \emph{\bibinfo{title}{How to compute using quantum walks}},
  \doi{10.24350/CIRM.V.19600203}.
\newblock \bibinfo{note}{CIRM. Audiovisual resource.}

\bibitemdeclare{misc}{Lida2019}
\bibitem{Lida2019}
\bibinfo{author}{Daniel \surnamestart Lidar\surnameend} (\bibinfo{year}{2019}):
  \emph{\bibinfo{title}{Arbitrary-time error suppression for Markovian
  adiabatic quantum computing using stabilizer subspace codes}}.
\newblock \urlprefix\url{http://arxiv.org/abs/1904.12028}.

\bibitemdeclare{misc}{Lode2019}
\bibitem{Lode2019}
\bibinfo{author}{Bas \surnamestart Lodewijks\surnameend}
  (\bibinfo{year}{2019}): \emph{\bibinfo{title}{Mapping NP-hard and NP-complete
  optimisation problems to Quadratic Unconstrained Binary Optimisation
  problems}}.
\newblock \urlprefix\url{http://arxiv.org/abs/1911.08043}.

\bibitemdeclare{article}{Love2010}
\bibitem{Love2010}
\bibinfo{author}{N.~B. \surnamestart Lovett\surnameend},
  \bibinfo{author}{S.~\surnamestart Cooper\surnameend}, \bibinfo{author}{M.~S.
  \surnamestart Everitt\surnameend}, \bibinfo{author}{M.~\surnamestart
  Trevers\surnameend} \& \bibinfo{author}{V.~\surnamestart Kendon\surnameend}
  (\bibinfo{year}{2010}): \emph{\bibinfo{title}{Universal quantum computation
  using the discrete time quantum walk}}.
\newblock {\sl \bibinfo{journal}{Phys.~Rev.~A}} \bibinfo{volume}{81}, p.
  \bibinfo{pages}{042330}, \doi{10.1103/PhysRevA.81.042330}.

\bibitemdeclare{article}{Mohs2008}
\bibitem{Mohs2008}
\bibinfo{author}{M.~\surnamestart Mohseni\surnameend},
  \bibinfo{author}{P.~\surnamestart Rebentrost\surnameend},
  \bibinfo{author}{S.~\surnamestart Lloyd\surnameend} \&
  \bibinfo{author}{A.~\surnamestart Aspuru-Guzik\surnameend}
  (\bibinfo{year}{2008}): \emph{\bibinfo{title}{Environment-assisted quantum
  walks in photosynthetic energy transfer}}.
\newblock {\sl \bibinfo{journal}{J.~Chem.~Phys.}} \bibinfo{volume}{129}, p.
  \bibinfo{pages}{174106}, \doi{10.1063/1.3002335}.

\bibitemdeclare{article}{Mont2015}
\bibitem{Mont2015}
\bibinfo{author}{Ashley \surnamestart Montanaro\surnameend}
  (\bibinfo{year}{2018}): \emph{\bibinfo{title}{Quantum-Walk Speedup of
  Backtracking Algorithms}}.
\newblock {\sl \bibinfo{journal}{Theory of Computing}}
  \bibinfo{volume}{14}(\bibinfo{number}{15}), pp. \bibinfo{pages}{1--24},
  \doi{10.4086/toc.2018.v014a015}.

\bibitemdeclare{misc}{Mont2019}
\bibitem{Mont2019}
\bibinfo{author}{Ashley \surnamestart Montanaro\surnameend}
  (\bibinfo{year}{2019}): \emph{\bibinfo{title}{Quantum speedup of
  branch-and-bound algorithms}}.
\newblock \urlprefix\url{http://arxiv.org/abs/1906.10375}.
\newblock \bibinfo{note}{ArXiv:1906.10375}.

\bibitemdeclare{article}{Morl2019}
\bibitem{Morl2019}
\bibinfo{author}{JG~\surnamestart Morley\surnameend},
  \bibinfo{author}{N~\surnamestart Chancellor\surnameend},
  \bibinfo{author}{S~\surnamestart Bose\surnameend} \&
  \bibinfo{author}{V~\surnamestart Kendon\surnameend} (\bibinfo{year}{2019}):
  \emph{\bibinfo{title}{Quantum search with hybrid adiabatic-quantum walk
  algorithms and realistic noise}}.
\newblock {\sl \bibinfo{journal}{Phys.~Rev.~A}} \bibinfo{volume}{99}, p.
  \bibinfo{pages}{022339}, \doi{10.1103/PhysRevA.99.022339}.

\bibitemdeclare{article}{Pere2008}
\bibitem{Pere2008}
\bibinfo{author}{Hagai~B. \surnamestart Perets\surnameend},
  \bibinfo{author}{Yoav \surnamestart Lahini\surnameend},
  \bibinfo{author}{Francesca \surnamestart Pozzi\surnameend},
  \bibinfo{author}{Marc \surnamestart Sorel\surnameend},
  \bibinfo{author}{Roberto \surnamestart Morandotti\surnameend} \&
  \bibinfo{author}{Yaron \surnamestart Silberberg\surnameend}
  (\bibinfo{year}{2008}): \emph{\bibinfo{title}{Realization of quantum walks
  with negligible decoherence in waveguide lattices}}.
\newblock {\sl \bibinfo{journal}{Phys. Rev. Lett.}} \bibinfo{volume}{100}, p.
  \bibinfo{pages}{170506}, \doi{10.1103/PhysRevLett.100.170506}.

\bibitemdeclare{incollection}{Picc2017}
\bibitem{Picc2017}
\bibinfo{author}{Gualtiero \surnamestart Piccinini\surnameend}
  (\bibinfo{year}{2017}): \emph{\bibinfo{title}{Computation in Physical
  Systems}}.
\newblock In \bibinfo{editor}{Edward~N. \surnamestart Zalta\surnameend},
  editor: {\sl \bibinfo{booktitle}{The Stanford Encyclopedia of Philosophy}},
  \bibinfo{edition}{{S}ummer 2017} edition, \bibinfo{publisher}{Stanford
  University Press}.
\newblock
  \urlprefix\url{http://plato.stanford.edu/archives/sum2017/entries/computation-physicalsystems/}.

\bibitemdeclare{book}{Putn1988}
\bibitem{Putn1988}
\bibinfo{author}{Hilary \surnamestart Putnam\surnameend}
  (\bibinfo{year}{1988}): \emph{\bibinfo{title}{Representation and Reality}}.
\newblock \bibinfo{publisher}{MIT Press}, \bibinfo{address}{Cambridge, MA}.
\newblock
  \urlprefix\url{https://mitpress.mit.edu/books/representation-and-reality}.
\newblock \bibinfo{note}{{ISBN}: 9780262161084}.

\bibitemdeclare{article}{Rola2002}
\bibitem{Rola2002}
\bibinfo{author}{J\'er\'emie \surnamestart Roland\surnameend} \&
  \bibinfo{author}{Nicolas~J. \surnamestart Cerf\surnameend}
  (\bibinfo{year}{2002}): \emph{\bibinfo{title}{Quantum search by local
  adiabatic evolution}}.
\newblock {\sl \bibinfo{journal}{Phys.~Rev.~A}} \bibinfo{volume}{65}, p.
  \bibinfo{pages}{042308}, \doi{10.1103/PhysRevA.65.042308}.

\bibitemdeclare{article}{Ryan2005}
\bibitem{Ryan2005}
\bibinfo{author}{C.~A. \surnamestart Ryan\surnameend},
  \bibinfo{author}{M.~\surnamestart Laforest\surnameend},
  \bibinfo{author}{J.~C. \surnamestart Boileau\surnameend} \&
  \bibinfo{author}{R.~\surnamestart Laflamme\surnameend}
  (\bibinfo{year}{2005}): \emph{\bibinfo{title}{Experimental implementation of
  discrete time quantum random walk on an {NMR} quantum information
  processor}}.
\newblock {\sl \bibinfo{journal}{Phys.~Rev.~A}} \bibinfo{volume}{72}, p.
  \bibinfo{pages}{062317}, \doi{10.1103/PhysRevA.72.062317}.

\bibitemdeclare{article}{Schr2011}
\bibitem{Schr2011}
\bibinfo{author}{A.~\surnamestart Schreiber\surnameend}, \bibinfo{author}{K.~N.
  \surnamestart Cassemiro\surnameend}, \bibinfo{author}{V.~\surnamestart
  Poto\v{c}ek\surnameend}, \bibinfo{author}{A.~\surnamestart
  G\'{a}bris\surnameend}, \bibinfo{author}{I.~\surnamestart Jex\surnameend} \&
  \bibinfo{author}{Ch. \surnamestart Silberhorn\surnameend}
  (\bibinfo{year}{2011}): \emph{\bibinfo{title}{Decoherence and disorder in
  quantum walks: From ballistic spread to localization}}.
\newblock {\sl \bibinfo{journal}{Phys.~Rev.~Lett.}} \bibinfo{volume}{106}, p.
  \bibinfo{pages}{180403}, \doi{10.1103/PhysRevLett.106.180403}.

\bibitemdeclare{article}{Shen2002}
\bibitem{Shen2002}
\bibinfo{author}{Neil \surnamestart Shenvi\surnameend}, \bibinfo{author}{Julia
  \surnamestart Kempe\surnameend} \& \bibinfo{author}{K~\surnamestart {Birgitta
  Whaley}\surnameend} (\bibinfo{year}{2003}): \emph{\bibinfo{title}{A quantum
  random walk search algorithm}}.
\newblock {\sl \bibinfo{journal}{Phys.~Rev.~A}} \bibinfo{volume}{67}, p.
  \bibinfo{pages}{052307}, \doi{10.1103/PhysRevA.67.052307}.

\bibitemdeclare{article}{Steff2003}
\bibitem{Steff2003}
\bibinfo{author}{\surnamestart Steffen\surnameend},
  \bibinfo{author}{\surnamestart vanDam\surnameend},
  \bibinfo{author}{\surnamestart Hogg\surnameend},
  \bibinfo{author}{\surnamestart Breyta\surnameend} \&
  \bibinfo{author}{\surnamestart Chuang\surnameend} (\bibinfo{year}{2003}):
  \emph{\bibinfo{title}{Experimental implementation of an adiabatic quantum
  optimization algorithm}}.
\newblock {\sl \bibinfo{journal}{Phys.~Rev.~Lett.}}
  \bibinfo{volume}{90}(\bibinfo{number}{6}), p. \bibinfo{pages}{067903},
  \doi{10.1103/PhysRevLett.90.067903}.

\bibitemdeclare{inproceedings}{Step2019}
\bibitem{Step2019}
\bibinfo{author}{S.~\surnamestart Stepney\surnameend} \&
  \bibinfo{author}{V.~\surnamestart Kendon\surnameend} (\bibinfo{year}{2019}):
  \emph{\bibinfo{title}{The role of the representational entity in physical
  computing}}.
\newblock In: {\sl \bibinfo{booktitle}{UCNC 2019, Tokyo, Japan, June 2019}},
  {\sl \bibinfo{series}{LNCS}} \bibinfo{volume}{11493},
  \bibinfo{publisher}{Springer}, pp. \bibinfo{pages}{219--231},
  \doi{10.1007/978-3-030-19311-9_18}.

\bibitemdeclare{incollection}{Wies2008}
\bibitem{Wies2008}
\bibinfo{author}{K.~\surnamestart {Wiesner}\surnameend} (\bibinfo{year}{2009}):
  \emph{\bibinfo{title}{{Quantum Cellular Automata}}}.
\newblock In \bibinfo{editor}{Robert~A. \surnamestart Meyers\surnameend},
  editor: {\sl \bibinfo{booktitle}{Springer Encyclopedia of Complexity and
  System Science}}, chapter \bibinfo{chapter}{Cellular Automata, Mathematical
  Basis of, Ed.~Andy Adamatzky}, \bibinfo{publisher}{Springer}, p.
  \bibinfo{pages}{00105}, \doi{10.1007/978-0-387-30440-3_426}.
\newblock \bibinfo{note}{{o}r http://arxiv.org/abs/0808.0679}.

\end{thebibliography}
